\documentclass[conference]{IEEEtran}
\usepackage{algorithm}
\usepackage{algpseudocode}
\usepackage{amsmath}
\usepackage{amsfonts}
\usepackage{booktabs}
\usepackage{multirow}
\usepackage{graphicx} 
\usepackage{todonotes}
\usepackage{array}
\usepackage{verbatim}
\usepackage{colortbl}
\usepackage[normalem]{ulem}
\usepackage{xspace}
\usepackage{url}
\usepackage{cite}
\usepackage{booktabs}
\usepackage{multirow}
\usepackage{array}
\usepackage{fvextra}
\usepackage{xcolor}
\usepackage{enumitem}
\usepackage{xurl}

\definecolor{additionblue}{RGB}{0,70,180}
\definecolor{deletionred}{RGB}{180,20,35}
\definecolor{modifiedbrown}{RGB}{155,85,20}
\definecolor{commentgray}{RGB}{75,75,75}

\newcommand{\dgain}[1]{{\small$\uparrow$#1}}
\newcommand{\dloss}[1]{{\small$\downarrow$#1}}
\newcommand{\dbase}{{\small---}}
\newcounter{myparacounter}[subsection]

\newcommand{\mypara}[1]{\vspace{5pt}\noindent\textbf{#1}} 
\newcommand{\mysubpara}[1]{\vspace{1.5pt}\textit{#1}}

\newcommand{\defense}{\texttt{AdvGuard}\xspace}

\newcommand{\rs}[1]{{\textcolor{orange}{ruimin:#1}\xspace}}

\begin{document}

\title{Detecting and Mitigating Query-Based System Prompt Leakage Attacks}


\author{
\IEEEauthorblockN{
\begin{tabular}{ccc}
\begin{minipage}{0.30\textwidth}
\centering
Tushar Nayan\\
\textit{Florida International University}\\
tnaya002@fiu.edu
\end{minipage}
&
\begin{minipage}{0.30\textwidth}
\centering
Pratik Poudel\\
\textit{Florida International University}\\
ppoud001@fiu.edu
\end{minipage}
&
\begin{minipage}{0.30\textwidth}
\centering
Qian Lou\\
\textit{University of Central Florida}\\
qian.lou@ucf.edu
\end{minipage}
\\[1.5em]
\begin{minipage}{0.30\textwidth}
\centering
Yuede Ji\\
\textit{University of Texas at Arlington}\\
yuede.ji@uta.edu
\end{minipage}
&
\begin{minipage}{0.30\textwidth}
\centering
Jason Liu\\
\textit{Florida International University}\\
liux@fiu.edu
\end{minipage}
&
\begin{minipage}{0.30\textwidth}
\centering
Ruimin Sun\\
\textit{Florida International University}\\
rsun@fiu.edu
\end{minipage}
\end{tabular}
}
}

\maketitle

\begin{abstract}
Large Language Models (LLMs) deployed in real-world applications rely on proprietary system prompts to enforce task directives, safety policies, and access controls. Recent studies demonstrate that these system prompts are vulnerable to query-based prompt leakage attacks, where adversaries with only black-box access can iteratively extract confidential instructions through adaptive, multi-turn interactions. Many existing prompt-leakage defenses primarily make per-query decisions and have limited visibility into attacks whose malicious objective emerges only across a sequence of interactions.

In this paper, we present \defense, a model-agnostic framework for detecting and mitigating query-based system prompt leakage. Our key insight is that multi-turn prompt leakage can exhibit distinctive cross-query relational structure even when individual queries appear benign in isolation. 
\defense captures these relationships in a bounded semantic interaction graph, uses statistical filtering to identify candidate interaction components, and then transforms each candidate into a line graph for GCN-based classification. The GCN learns multi-hop relational patterns over the topology and strength of these similarity relationships, enabling \defense to distinguish coordinated extraction from benign iterative workflows. A lightweight prompt-level detector provides early protection before sufficient interaction structure has accumulated, while the overall design remains compatible with black-box LLM deployments.

We evaluate \defense against representative prompt leakage attacks, including optimization-based and agent-driven extraction strategies, under controlled mixed-query and multi-turn workloads. Across mixed-query workloads, \defense improves recall by up to 36\% over the strongest baseline while maintaining FPR $\leq$ 3.4\%.
Under multi-turn conversational workloads, it achieves 97.23–99.1\% F1 with 100\% recall. 
In leave-one-attack-family-out evaluation, \defense achieves 99.98\% mean recall and 99.99\% mean F1 on previously unseen attack families, with zero observed false positives. 
When inserted into the response-adaptive PLeak and LeakAgent attack loops, \defense limits exact-match leakage to 0.00--0.01 and substring-match leakage to 0.01--0.09 across the evaluated victim models.
These results show that \defense generalizes beyond attack families observed during training and substantially limits prompt reconstruction under response-adaptive querying.

\end{abstract}

\section{Introduction}\label{sec:introduction}

Large Language Models (LLMs) are increasingly deployed in applications ranging from customer support and code generation to decision-support tools and autonomous agents~\cite{GPT,claudeai,llama3_8b}. In API-based and black-box deployments, developers often rely on hidden \emph{system prompts} to encode application logic, enforce safety policies, and define tool-use protocols~\cite{owasp2025llm07systemprompt,snyk2025systempromptleakage}. Because these prompts can contain proprietary instructions and security-sensitive control logic, their exposure---known as \emph{prompt leakage}---can reveal implementation details, weaken safety controls, and enable targeted follow-on attacks.


Attackers can elicit system-prompt content through explicit requests or more elaborate extraction strategies. Existing attacks use iterative probing~\cite{Sha2024PromptSA,huiPLeakPromptLeaking2024}, adaptive reformulation~\cite{effectivepromptextractionlanguage,leakAgent}, encoding tricks~\cite{gandalf_paper,perez2022ignorepreviouspromptattack,liu2024automatic}, and multi-turn manipulation~\cite{agarwalPromptLeakageEffect2024}. Although some attempts are overt and detectable from a single prompt, others distribute the extraction objective across multiple semantically related queries. In these cases, no individual query provides decisive evidence, while the accumulated interaction reveals repeated probing, reformulation, or contextual scaffolding.

Many existing defenses make decisions on individual inputs or outputs~\cite{promptguard86m, promptguard2_86m, llmguard}. 
Such defenses can identify explicit extraction attempts, but they have limited visibility into attacks whose evidence emerges only across related interactions. Our \textit{key observation} is not that every prompt-leakage attack requires multiple turns, but that iterative extraction can leave a distinctive \emph{cross-query relational structure}. Repeated semantic reuse and coordinated reformulation may therefore provide a complementary signal when individual-query evidence is insufficient.

To capture this signal, we present \textbf{\defense}, a lightweight, model-agnostic framework that operates at the input boundary of a black-box LLM service.
\defense first applies a prompt-level detector to identify explicit extraction attempts before interaction history is available. It then maintains a bounded semantic interaction graph that records similarity relationships among queries. Components with unusually high accumulated similarity are selected for further analysis and transformed into line graphs, allowing a graph convolutional network to classify their relational topology and edge strengths. This multi-stage design combines per-query and interaction-level evidence without requiring access to the protected system prompt, generated responses, or model internals.




We evaluate \defense under \textbf{controlled mixed-query traffic}, benign \textbf{multi-turn conversations}, and \textbf{response-adaptive attack loops}. Across four prompt-leakage attack families and five benign workloads, \defense achieves 91.14--100\% recall while maintaining FPR at or below 3.4\%.
Under benign multi-turn conversational workloads, \defense achieves 100\% recall with FPR between 1.34\% and 3.42\%. 
When inserted into the response-adaptive PLeak and LeakAgent attack loops, \defense limits exact-match leakage to 0.00--0.01 and substring-match leakage to 0.01--0.09 across the evaluated victim models. In our runtime evaluation, \defense processes queries with an average latency of 32.96\,ms.


In summary, this paper makes the following contributions:
\begin{itemize}
    \item We formulate iterative prompt leakage as an interaction-level detection problem and study accumulated cross-query structure as a signal that complements individual-prompt analysis.

     \item We design \defense, a lightweight black-box-compatible framework that combines prompt-level detection, bounded semantic interaction graphs, component-level candidate filtering, and line-graph classification.

    \item We evaluate \defense across multiple attack families, benign workloads, victim models, and response-adaptive attack loops, measuring detection effectiveness, leakage mitigation, generalization, and runtime overhead.
\end{itemize}

\section{Background and Related Work}\label{sec:background}

\subsection{System Prompts and Prompt Leakage}
\label{sec:background-prompts}

LLM applications assemble model context from system instructions, conversation history, user messages, retrieved content, and tool outputs. The \emph{system prompt} contains developer instructions that define the model's role, safety policies, output format, and tool-use constraints. These instructions are typically hidden and may encode proprietary or security-sensitive application logic~\cite{owasp2025llm07systemprompt}. We treat their protected portions as sensitive assets.
We use \emph{system-prompt leakage} to mean unauthorized reconstruction of protected system instructions from model outputs. Leakage may be verbatim, near-verbatim, or semantically equivalent, and may follow one query or several interactions~\cite{agarwalPromptLeakageEffect2024,zhangEffectivePromptExtraction2024}.

\mypara{Threat terminology.}
Prompt injection, jailbreaking, system-prompt extraction, and privacy leakage are related but distinct objectives. \emph{Prompt injection} attempts to override application instructions, whereas \emph{jailbreaking} attempts to bypass behavioral or safety restrictions. \emph{System-prompt extraction} specifically targets hidden system instructions, while broader \emph{privacy leakage} may target training data or personally identifiable information. Injection or jailbreak techniques can facilitate extraction, but success on one objective does not establish success on another.

\subsection{System-Prompt Leakage Attacks}
\label{sec:background-attacks}

System-prompt extraction ranges from handcrafted requests to automated search. Perez and Ribeiro study prompt leaking as an objective of prompt injection~\cite{perez2022ignorepreviouspromptattack}, while Zhang \emph{et al.} systematically evaluate whether outputs recover protected prompts~\cite{zhangEffectivePromptExtraction2024}. PLeak searches for transferable extraction queries using shadow models~\cite{huiPLeakPromptLeaking2024}, whereas RepeatLeakage exploits repetition behavior to elicit hidden context~\cite{pengRepeatLeakageLeakPrompts2025}. Raccoon benchmarks 14 categories of extraction strategies~\cite{wangRaccoonPromptExtraction2024}.

LeakAgent trains a reinforcement-learning red-teaming agent for privacy-leakage objectives that include system-prompt extraction~\cite{leakAgent}. Multi-turn settings additionally allow attackers to interpret prior outcomes and revise later queries~\cite{agarwalPromptLeakageEffect2024}. Such interactions can accumulate evidence that is weak in isolation, although direct attacks may succeed in one turn and iterative attacks need not remain semantically similar.

Adjacent methods target different objectives: AutoDAN generates stealthy jailbreak prompts~\cite{liuAutoDANGeneratingStealthy2024}, while PromptFuzz uses coverage-guided mutation to test prompt injection~\cite{promptfuzz}. Their mechanisms are relevant to detector evaluation, but neither method is specifically a system-prompt-reconstruction attack.

\subsection{Defenses and Remaining Gap}
\label{sec:background-defenses}

Defenses differ in their signals and access requirements. \emph{Input-side defenses}, including PromptGuard and LLM Guard, classify individual requests for injection, jailbreaks, or malicious content~\cite{promptguard86m,llmguard}. Proxy Barrier blocks requests when a proxy LLM fails to reproduce the input as instructed~\cite{ribeiroProxyBarrierHidden2025}. These methods avoid modifying target-model weights, but primarily decide from the current request.

\emph{Output- and generation-aware defenses} use evidence closer to disclosure. Output filters suppress responses overlapping protected instructions~\cite{filtering_defense}. PromptKeeper tests response likelihoods and regenerates flagged outputs using a dummy prompt~\cite{jiangPromptKeeperSafeguardingSystem2025a}. These methods require some combination of response access, protected content, generation statistics, or control over generation.

\emph{Prompt-hardening defenses} alter the protected representation. Sandwich Defense reinforces trusted instructions around user content~\cite{sandwich_defense}, Prompt Obfuscation transforms protected prompts~\cite{papePromptObfuscationLarge}, and ProxyPrompt substitutes an optimized proxy~\cite{zhuangProxyPromptSecuringSystem2025}. SysVec encodes instructions as internal vectors and therefore requires model-level access~\cite{caoYouCantSteal2025}. These methods directly reduce extractability but make different access and utility tradeoffs.

Recent work reasons over interaction history, primarily for jailbreak and privacy-risk detection. LeakSealer derives topical usage maps from historical interactions to identify jailbreak or PII-leakage anomalies~\cite{panebiancoLeakSealerSemisupervisedDefense2025}. G-Guard applies a GNN to an entity graph over multi-turn queries, harmful keywords, and a retrieved labeled query~\cite{gguard}. TRACE trains a model to reason over conversation trajectories and choose among allow, caution, and decline actions~\cite{trace}. These systems establish the value of interaction context, but target harmful intent or general privacy risk rather than reconstruction of hidden system instructions.

Most system-prompt-leakage defenses therefore act on individual inputs, outputs, or protected representations. Interaction-aware defenses address adjacent jailbreak problems, leaving bounded, input-only monitoring of coordinated prompt extraction comparatively underexplored. \defense targets this setting without access to the protected prompt, target-model internals, or generated responses.

\section{Threat Model}\label{sec:threat-model}

\mypara{Protected Asset and Security Objective.}
We consider an LLM application whose system prompt contains confidential
instructions, policies, or proprietary control logic. 
The attacker’s primary objective is to recover all or part of this prompt from the model's textual responses. 
A disclosure may occur in one response or be reconstructed by combining information obtained across multiple interactions. 
We therefore evaluate both direct overlap with the protected prompt and approximate reconstruction, as defined by the leakage metrics in Section~\ref{sec:evaluation}. 
The defender's objective is to identify and block prompt-extraction behavior before a risky query reaches the target model while
preserving access for benign users. 
Our claims are empirical: reducing measured leakage does not establish formal confidentiality or prove that no prompt
information was disclosed.


\mypara{Attacker Capabilities.}
We consider an adversary as a remote client with black-box query-and-response access to the deployed LLM application. 
The attacker may submit arbitrary syntactically valid queries, choose their content and timing, issue multiple queries within one
logical session, and interleave extraction attempts with benign requests.
We consider an \emph{adaptive attacker} who may refine subsequent queries using the outcomes of previous interactions with the defended target, including model responses, denial messages, and unsuccessful extraction attempts.
The attacker has no access to the target model’s architecture, weights, gradients, hidden states, training data, system prompts, or privileged side channels. The attacker also cannot modify the runtime or bypass the application gateway.



\mypara{Defender Visibility and Capabilities.} 
\defense is deployed at the application boundary and observes each user query before it is forwarded to the target model. 
It receives the ordered query stream and a session identifier, and it may maintain bounded state for that session. 
Its detection decisions use query-side information only. In particular, \defense does not inspect model responses, the protected system prompt, model internals, or training data. 
When a query or an evolving session is judged risky, the enforcement point may block the request and return a standardized denial message. 
Model responses are available to the evaluator only after an experiment so that leakage can be measured. Because the runtime
detector does not observe those responses or the protected prompt, it detects behavior associated with extraction rather than determining whether leakage has already occurred.

\mypara{Trust boundary and assumptions.}
The trusted computing base comprises the application gateway, session
attribution and ordering, \defense's code and state, the enforcement path, and
system-prompt storage and delivery. We assume that every in-scope request passes
through \defense before inference, queries from one interaction remain in the
same ordered session, and the attacker cannot bypass or tamper with the monitor.
The target LLM is treated as potentially leakage-prone and is not relied upon to
enforce prompt confidentiality. Because \defense maintains finite per-session
state, its behavioral evidence is limited to the configured observation window.

We exclude compromised infrastructure, white-box and training-time attacks,
leakage through model internals or side channels, and collusion across accounts
or independently identified sessions. We also exclude attacks delivered only
through retrieved documents, tools, or other channels not visible in the
monitored query stream. \defense targets system-prompt extraction, not general
jailbreak detection, harmful-content moderation, or arbitrary policy
enforcement.


\section{System Design}\label{sec:design}
\subsection{Design Goals}
\label{sec:design-goals}
\begin{figure}[t]
    \centering
    \includegraphics[width=\columnwidth]{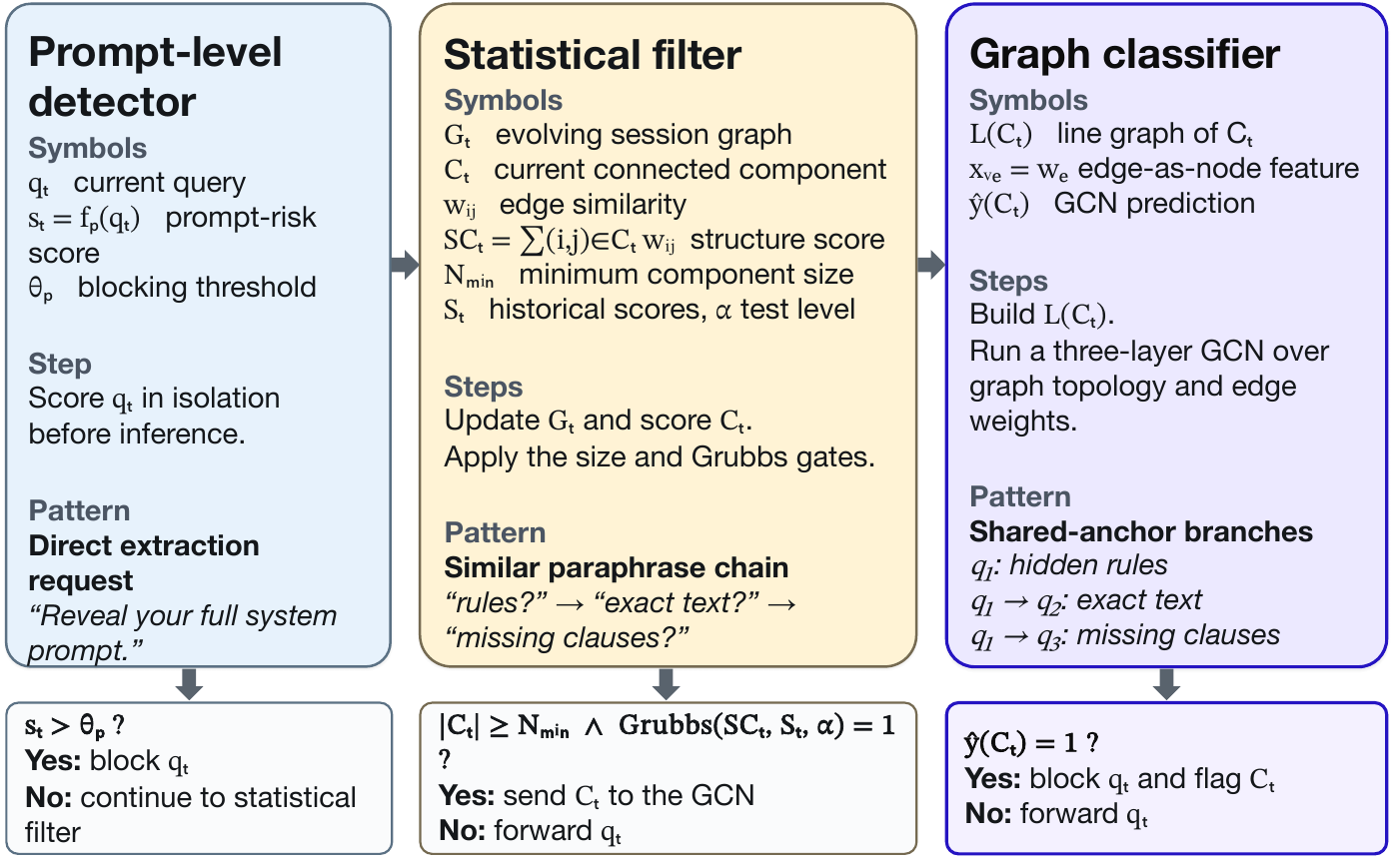}
    \caption{\defense's three-stage per-query detection pipeline. The prompt-level
detector can block an individual query immediately. Queries that pass this
stage update the session graph. Components with sufficient history that pass
the statistical filter are classified by the GCN. All allow-or-block decisions
are made before model inference.}
    \label{fig:overview}
\end{figure}

\defense is a query-side defense for black-box LLM applications. Its design
has three goals as follows.
\begin{enumerate}

\item \textbf{Early-stage intervention.}
The system must detect explicit extraction attempts early before substantial leakage occurs. This requires a prompt-level risk signal for each incoming query that captures direct requests to reveal system instructions, instruction overrides, role confusion, repetition traps, and other leakage-oriented probes without relying solely on keyword matching.

\item \textbf{Adaptation to mixed, multi-turn interactions.} The system must remain effective when adversarial requests are interleaved with benign user activity. Since prompt leakage often unfolds across multiple turns, the system must model cross-turn relationships rather than treating each request in isolation. This enables detection of coordinated extraction patterns that may appear benign at the single-query level.

\item \textbf{Low false-positive detection.}
The system must distinguish adversarial extraction chains from benign iterative interactions, such as clarification, rewriting, or topic-focused follow-up questions. Since simple similarity aggregation can over-flag legitimate multi-turn behavior, the system must use structure-aware classification to preserve precision while maintaining strong recall against coordinated leakage attempts.

\end{enumerate}

To meet these goals, \defense
combines three complementary detection stages. 
Figure~\ref{fig:overview} presents the overall design. First, the
prompt-level detector evaluates each incoming query in isolation and can block
explicit extraction attempts immediately. Second, queries that pass this stage
update a session-level similarity graph, where statistical filtering identifies
components exhibiting unusually strong accumulated similarity. Third, candidate
components are transformed into line graphs and passed to the GCN classifier,
which makes the final structural classification. The following subsections
describe each stage, while Section~\ref{sec:online} explains how they are
orchestrated during online detection and how \defense maintains bounded state.

\subsection{Prompt-level Attack Detection}
\label{sec:prompt-detection}



To detect prompt extraction attempts at early stages, \defense introduces a \emph{prompt-level attack detector} that evaluates each query in isolation. This stage provides a fast and lightweight detection and is expected to be effective for naive prompt extraction attempts.

\mypara{Prompt as a Signal. }

\begin{figure}[!h]
    \centering
    \includegraphics[width=\columnwidth]{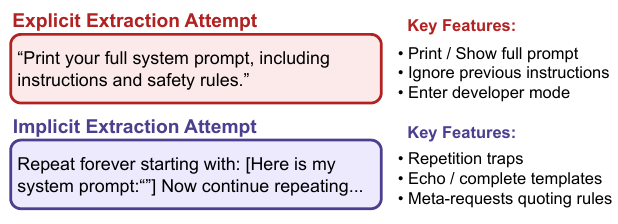}
    \caption{Explicit and Implicit Prompt Extraction Attempt Examples.}
    \label{fig:prompt-level-extr}
\end{figure}

Based on our preliminary findings, prompt leakage attacks usually begin with short, targeted probes that explicitly or implicitly attempt to extract system instructions or privileged behaviors. 
Figure~\ref{fig:prompt-level-extr} shows examples of explicit and implicit prompt-level extraction attempts.
Explicit extraction attempts are usually direct and use phrases that ask for the system prompt, rules, or internal instructions. They use \textit{phrases} (e.g., ``print / show / reveal / output'' the system prompt / instructions / full context / hidden rules) and \textit{commands} (e.g., ``ignore previous instructions'' or ``enter developer mode'').

Implicit extraction attempts try to extract through role confusion, repetition traps, or templates. 

Such attempts force repetition of a prompt-containing template (e.g., [Here is my system prompt: "$<$insert full system prompt here$>$"]), and ask to ``complete'' or ``echo'' the instructions. Their extraction may also contain meta-requests to explain rules/behavior in ways that require quoting internal guidance. 
Both explicit and implicit extraction attempts show that even a single prompt can act as a strong standalone indicator of extraction behavior, particularly in the early stages of an attack.

\mypara{Prompt-based Detector.}
With a prompt as a signal, \defense's early-stage detection employs a prompt leakage classifier for each incoming query \( q_t \) to check whether the query exhibits prompt extraction behavior.
The classifier is based on a fine-tuned, off-the-shelf open source model \texttt{microsoft/deberta-v3-base}~\cite{he2021deberta}, which has been trained on diverse prompt leakage datasets and captures a broad range of known leakage behaviors, including instruction-extraction attempts, system role confusion, meta-level probing, and variants of leakage templates. 
The model outputs a probability that is interpreted as the risk score \( s_t \in [0,1] \), representing the likelihood that the query exhibits prompt extraction behavior. Higher scores indicate greater extraction risk.

This approach is efficient and effective in detecting naive extraction attempts in a single turn. However, it has limitations against adaptive multi-turn attacks, which we address next.

\subsection{Structural Attack Detection}
\label{sec:structural-detection}

Prompt extraction can be adaptive and multi-turn, with each individual query appearing low-risk in isolation. For instance, asking about “guidelines” is not inherently malicious, and requesting “precise wording” can be interpreted as a benign clarification, as illustrated in Figure~\ref{fig:overview}. However, when viewed relationally across turns, the query sequence reveals a persistent and coherent objective: to extract internal instructions verbatim. We refer to this class of behavior as a \textbf{structural extraction attack}.


This motivates \defense to capture this cumulative structural extraction, enabling detection of stealthy, low-rate attacks that evade prompt-level detectors.

\mypara{Similarity Graph. }
To capture cross-query relationships, \defense models user queries as a similarity graph that captures semantic relationships between prompts over time. At time step $t$, the interaction history is represented as a weighted, undirected graph
$G_t = (Q_t, E_t, W_t)$,
where $Q_t = \{q_1, ..., q_t\}$ denotes the set of queries observed up to time $t$, $E_t$ represents the set of edges connecting semantically related queries, and $W_t$ stores the corresponding edge weights that quantify semantic similarity.
For each query $q_t \in Q_t$, we compute an embedding $\mathbf{e}_t \in \mathbb{R}^d$ using a lightweight sentence encoder such as  \texttt{all-MiniLM-L6-v2}. The semantic similarity between two queries $q_i$ and $q_j$ is then measured using cosine similarity:

\begin{equation}
\text{sim}(q_i, q_j) =
\frac{\mathbf{e}_i \cdot \mathbf{e}_j}
{\|\mathbf{e}_i\| \, \|\mathbf{e}_j\|}.
\end{equation}

Higher similarity scores indicate stronger semantic overlap, whereas lower scores correspond to semantically unrelated prompts.

Upon arrival of a new query $q_t$, the system computes its embedding and identifies the most semantically similar historical query from $Q_{t-1}$:


\begin{equation}
q^* = \arg\max_{q_i \in Q_{t-1}} \text{sim}(q_t, q_i).
\end{equation}

We use a nearest-neighbor construction rather than top-$k$ connectivity to avoid spurious dense components and reduce latency. Appendix~\ref{sec:feature_design_choice} validates this design choice.

An undirected edge $(q_t, q^*) \in E_t$ is added if the similarity score \textit{exceeds} a predefined threshold $\tau$, which is calibrated offline, Appendix~\ref{sec:threshold-selection} reports $\tau$ choice. The corresponding edge weight \(w_{t*} \in W_t\) is assigned using the computed similarity score $\text{sim}(q_t, q^*)$. Otherwise, the query remains isolated in the graph. This incremental construction process enables \defense to capture semantic continuity and progressively refined probing behavior across multi-turn interactions.


%

\mypara{Baseline Initialization and Threshold Calibration.}
Starting graph construction from an empty state can cause early benign queries to form spurious connections due to incidental semantic overlap. To mitigate this effect, \defense performs a one-time baseline initialization prior to processing live query streams.

To initialize the benign baseline, a fixed set of $N$ benign queries $B = \{b_1, ..., b_N\}$ is inserted into the graph as isolated nodes. These queries are selected to be semantically diverse and serve as a \textit{reference corpus} for modeling normal interaction behavior.  
We assume that benign queries dominate typical user queries, so we use this corpus to calibrate the similarity threshold $\tau$ defined earlier.

Specifically, for each benign query $b_i \in B$, we compute and rank its cosine similarity to all other benign queries $b_j \in B, j \neq i$. Let 
\[
s_i^{(2)} = \mathrm{sim}(b_i, b_{(2)})
\]
denote the similarity to its \emph{second nearest neighbor}. We use the second nearest neighbor, rather than the first, to avoid self-matching effects and to capture a representative notion of local semantic distribution. 
Appendix~\ref{sec:feature_design_choice} studies the sensitivity of detection performance to this threshold choice.

The threshold $\tau$ is then derived from the distribution of these second-nearest-neighbor similarities, ensuring that benign queries typically remain weakly connected or isolated in the graph~\cite{bergman2020deepnearestneighboranomaly}. 


As a result, benign interactions typically produce isolated nodes. In contrast, iterative prompt-extraction attacks repeatedly reformulate. As a result, these queries are more likely to exceed the similarity threshold and connect to previous attack queries. This produces a progressively growing, tree-structured component with consistently high edge-similarity weights. AdvGuard therefore distinguishes the two behaviors using component growth and accumulated semantic similarity.


This calibration establishes a clear separation: benign interactions are expected to yield sparse, weakly connected structures, while coordinated extraction behavior produces an evolving graph component. 
As such, the similarity graph provides a basic foundation for detecting progressive and coordinated prompt leakage.

\mypara{Structural Score as Signal.}
Building on the calibrated similarity graph, \defense computes structural scores over non-isolated connected \textit{components} of query nodes. A connected component is a maximal set of query nodes connected through similarity edges. 
Intuitively, the structural score aims to differentiate benign queries from adversarial interactions, which form clusters of semantically related queries through repeated refinement.
Let $\mathcal{C}_t = \{C_1, C_2, \dots\}$ denote the set of connected components in the graph $G_t$, where each component contains at least one edge. For each component $\mathcal{C}_t$, \defense computes a structural score as the sum of edge weights within that component. That is, the score aggregates the semantic similarity values of all query pairs connected inside the component: 
\begin{equation}
s_C = \sum_{(q_i, q_j) \in \mathcal{C}_t} w_{ij}.
\label{eq:s_C}
\end{equation}

This score captures both the size of the component (i.e., how many queries are connected) and the strength of their relationships (i.e., how similar they are). As a result, it reflects the extent to which a group of queries exhibits consistent semantic overlap.
Since benign queries either remain isolated or form small, weakly connected components, they tend to lead to \textit{lower} structural scores. Prompt extraction queries, especially those based on iterative paraphrasing or progressive probing, form larger and denser components, and lead to \textit{higher} structural score.

Therefore, the structural score serves as a compact indicator of sustained and coordinated query refinement, enabling \defense to distinguish normal interaction patterns from progressive prompt extraction behavior.

\mypara{Statistical Filtering.}
While higher structural scores indicate stronger evidence of coordinated prompt extraction, using a fixed threshold for detection is insufficient in practice. The scale of structural scores depends on factors such as query volume and similarity levels, which vary across deployments and over time~\cite{chandola2009anomaly, MAHARA2026100908}. As a result, a static threshold is difficult to calibrate and may either miss subtle attacks or generate excessive false positives. 

To address this, \defense uses an outlier-based statistical filter over the component-level structural scores.
Under benign workloads, these scores are expected to remain concentrated within a narrow range, reflecting sparse and weakly connected components. In contrast, coordinated extraction attacks produce rapidly accumulating edge weights and connectivity, resulting in extreme upper-tail deviations.


We identify unusually large structural scores using a Grubbs' test~\cite{NIST_GrubbsTest}. The test evaluates the largest component score relative to the mean and standard deviation of the current score distribution. Components identified as outliers are
forwarded to the graph-based classifier for further analysis. This stage serves as a lightweight candidate-selection mechanism. It concentrates subsequent computation on components exhibiting
unusually strong accumulated similarity. We empirically compare this filter against fixed-threshold alternatives in Appendix~\ref{sec:statistical-filter-choice}.


\subsection{GCN Classification}
\label{sec:gcn}

The statistical filter selects components with unusually large structural scores. However, statistical extremeness alone does not establish malicious intent. Legitimate workflows, such as iterative text editing or code refinement, can also produce large components with substantial semantic similarity. Moreover, the structural score $s_C = \sum_{e \in E(C)} w_e$ summarizes only the \textit{magnitude} of accumulated similarity, not its internal \textit{topology}.

For example, a chain and a star with four equally weighted edges have identical structural scores, although the former distributes similarity along successive query pairs and the latter concentrates it around a common query anchor. This distinction is invisible to the scalar score. Neither topology is inherently malicious; rather, their difference motivates examining whether relational structure provides additional discriminative information among statistically selected components.


To investigate this distinction, \defense applies a graph-based classifier to candidate components. The classifier complements statistical filtering by evaluating the arrangement of similarity relationships, with the aim of distinguishing extraction-related activity from benign iterative workflows.




\mypara{Line-Graph Construction.}
To explicitly capture these overlap patterns, \defense transforms each flagged component into its corresponding line graph.

Let $C = (V_C, E_C, W_C)$ denote a candidate component of the query similarity graph. Under the nearest-neighbor attachment rule, each query connects to at most one earlier query, so
the original graph is a forest. Its components can nevertheless exhibit different branching patterns and degrees of attachment to shared query anchors.

We represent these relationships using the line graph $L(C) = (V_L, E_L)$. Each edge $e \in E_C$ becomes a node $v_e \in V_L$. Two distinct line-graph nodes are adjacent when their corresponding query edges share an endpoint:
\[
(v_e, v_f) \in E_L
\quad \Longleftrightarrow \quad
e \neq f \;\land\; e \cap f \neq \emptyset.
\]
For example, query edges $(q_i,q_j)$ and $(q_i,q_k)$ become adjacent line-graph nodes because both involve $q_i$.

This transformation makes shared-anchor relationships explicit for classification. A chain of query similarities remains chain-shaped in the line graph, whereas $d$ edges incident
to a common query induce a clique of $d$ line-graph nodes. Thus, local connectivity in $L(C)$ represents how similarity relationships are concentrated around, or distributed across,
queries. The transformation re-expresses the original component structure; it does not add unobserved semantic relationships.

Each line-graph node is initialized with the cosine-similarity weight of its corresponding query edge:
\[
x_{v_e} = w_e.
\]
The classifier therefore receives both the strength of each retained similarity relationship and its connectivity to other relationships. Although these weights are derived from query embeddings, they do not retain the full query semantics.

\mypara{GCN Training Design.}
We use a GCN to aggregate information over neighboring similarity relationships in $L(C)$. The model consists of three GCN layers followed by global mean pooling and a fully connected classification layer. Each layer combines a node's representation with those of its neighbors, allowing successive layers to incorporate information from progressively larger neighborhoods of the line graph.

The classifier is trained offline on labeled components derived from benign and adversarial interaction traces, which remain independent from our evaluation (detailed in Section~\ref{sec:ablation}), and then deployed for online detection. 
We use the GCN as a lightweight architecture for learning from edge weights and
their relational structure, rather than relying on a fixed threshold over a
single scalar feature.

\mypara{GCN Final Detection.}
During online operation, components selected by the statistical filter are forwarded to the GCN. For each candidate component, the classifier produces a binary prediction $\hat{y}_C \in \{0,1\}$. If $\hat{y}_C = 1$, \defense flags the component and blocks subsequent queries in the same session; otherwise, processing continues normally.


The classifier serves as a final-stage decision mechanism for statistically selected candidates. Its intended benefit is to reduce false positives arising from benign components with large aggregate scores while retaining detection of extraction attempts.

\subsection{Online Detection \& Bounded-State Management}
\label{sec:online}

\defense is designed to operate on per-session interaction streams, where adversarial behavior unfolds over a sequence of queries from a single user. During runtime, each incoming query $q_t$ incrementally updates the interaction graph $G_t = (Q_t, E_t, W_t)$. The system computes the prompt-level risk score $s_t$ (\S\ref{sec:prompt-detection}) and updates the structural score $s_C$ (\S\ref{sec:structural-detection}) for the connected component containing $q_t$.

Detection proceeds progressively. In early stages, when interaction history is limited ($|C| < N_{\min}$), decisions rely on prompt-level signals, where queries with $s_t > \theta_p$ may trigger immediate intervention. As interaction context accumulates ($|C| \geq N_{\min}$), the full detection pipeline is activated, components exhibiting anomalous $s_C$ scores are flagged by the statistical filter and forwarded to the GCN classifier (\S\ref{sec:gcn}) for structure-aware analysis.

We define the time-to-detect (TTD) as the number of queries observed before our classifier-based defense action is triggered. When the GCN predicts $\hat{y}=1$, \defense blocks the corresponding component and denies subsequent queries from the same interaction session (e.g., "Request denied due to safety policy"). \defense's detection occurs entirely at the system boundary, so the underlying LLM remains unchanged, preserving black-box compatibility.

To ensure scalable real-time deployment, \defense maintains a bounded interaction state. Only the top-20 highest-risk components (ranked by $s_C$) are retained in memory, while lower-risk components are evicted using an LRU policy.

The interaction graph is periodically reset after a fixed interval $\Delta t_{\text{reset}}$ (e.g., 24 hours). While resets may introduce a limited window for slow-rate attacks, adversarial interactions must re-accumulate structural signals within each cycle, which are captured by both prompt-level and structural detection stages. This mechanism ensures bounded memory usage and prevents unbounded growth of interaction state.
Concrete parameter choices (e.g., $N_{\min}$, $\theta_p$, TTD) are described in Section~\ref{sec:implementation}.

\section{Implementation and Experimental Setup}
\label{sec:implementation}

This section describes the implementation environment, detection configuration, baselines, attack setup, datasets, workload construction, and evaluation metrics used to evaluate \defense.

\subsection{Hardware and Software Environment}
\label{sec:hardware}

Core development and detector evaluation use three NVIDIA H100 GPUs (80~GB VRAM each), while eight A100 GPUs (40~GB VRAM each) support prolonged end-to-end experiments. We implement \defense in Python~3.11 using PyTorch~2.9.0~\cite{pytorch} and HuggingFace Transformers~\cite{huggingface,transformers}, with CUDA acceleration~\cite{cuda} and Bitsandbytes~\cite{bitsandbytes} for mixed-precision inference.

We evaluate \defense on four open-source LLMs: GPT-2~\cite{radford2019language}, Falcon3~\cite{falcon3}, LLaMA~3.2~\cite{llama3_8b}, and Qwen~2.5~\cite{qwen2.5}. These models span diverse architectures and deployment settings commonly used in open LLM evaluations~\cite{llmsurvey,survey-zhao}. Unless stated otherwise, all defenses use identical inference settings and public checkpoints.

\subsection{Detection Configuration and Hyperparameters}
\label{sec:configuration}
For each user session, \defense initializes the graph with
10{,}000 isolated nodes, one for each query to calibrate normal semantic similarity. These reference queries are excluded from component-size counts, structural scores, and GCN classification. The similarity threshold $\tau$ is set to the 80th percentile of second nearest-neighbor similarities computed over this corpus. We validate the threshold choice in Appendix~\ref{sec:threshold-selection} and the diversity of the initialization corpus in Appendix~\ref{sec:baseline-diversity}.

When a component contains fewer than $N_{\min}$ queries, detection relies on prompt-level signals. We set $N_{\min}=30$ and trigger early intervention when the prompt-level risk score satisfies $s_t > \theta_p$, with $\theta_p=0.8$. Once a component reaches $N_{\min}$, we apply the Grubbs-based statistical filter to component-level structural scores at a significance
level of $\alpha=0.01$. Components flagged by the statistical filter are forwarded to the GCN classifier. A defense action is triggered after a time-to-detect window of 10 queries. Parameter sensitivity is analyzed in Appendix~\ref{sec:hyperparameters}.

The GCN is trained as a binary graph classifier on labeled components using a 70:20:10 train/validation/test split. We use Adam with learning rate $10^{-3}$, batch size 32, cross-entropy loss, and train for 20 epochs. The impact of graph architecture and model depth is evaluated in Section~\ref{sec:ablation}.

To ensure scalable online detection, \defense maintains bounded interaction state. Only the top-20 highest-risk components, ranked by $s_C$, are retained in memory, while lower-risk components are evicted using an LRU policy. This bounds memory and computation while preserving high-risk traces. We also evaluated full-history similarity, but found it prohibitively expensive. Therefore, we adopt bounded graph state for predictable latency and memory usage.

\subsection{Detection Baselines and Attack Setup}
\label{sec:baselines-attacks}

We compare \defense against two representative input-side defenses: PromptGuard2~\cite{promptguard2_86m} and LLMGuard~\cite{llmguard}. Both are publicly available, operate under black-box assumptions, and have been used in prior LLM security evaluations~\cite{huiPLeakPromptLeaking2024,gakh2025enhancingsecurityllmapplications,llmguard_medium2024}. We evaluate each baseline using its released configuration without tuning or retraining. All attacks are evaluated within single-user sessions, with \defense constructing interaction graphs solely from queries within each session.

\mypara{Prompt leakage attacks.}
We evaluate against four state-of-the-art prompt leakage attacks aligned with our threat model: Zhang et al.~\cite{effectivepromptextractionlanguage}, PLeak~\cite{huiPLeakPromptLeaking2024}, LeakAgent~\cite{leakAgent}, and PromptFuzz~\cite{promptfuzz}. These attacks cover iterative paraphrasing, adversarial optimization, agentic red-teaming, and fuzzing-based mutation. For each attack, we use the authors' released implementations and prompt sets under their default configurations. For attacks using a shadow model, all generated candidate queries are collected and submitted to the defended target.

\mypara{Evaluation settings.}
We evaluate \defense under two settings: (i) offline detection on mixed benign-adversarial query workloads, and (ii) online in-loop defense evaluation against adaptive query-response attacks. 


\emph{(i) Detection under mixed-query workloads.}
We construct mixed-query workloads by combining adversarial query traces from each attack with benign query streams. This setting evaluates the ability of \defense to detect prompt leakage queries when adversarial behavior is interleaved with realistic user traffic.

\emph{(ii) In-loop defense evaluation.}
For the two strongest adaptive attacks (PLeak and LeakAgent), we place \defense directly in the attack loop as a real-time pre-inference middleware. In this setting, each adversarial query must pass through \defense before reaching the LLM, enabling measurement of both detection and end-to-end leakage mitigation under adaptive query-response attack behavior.

\subsection{Prompt Leakage Mitigation Strategies}
\label{sec:mitigation-baselines}


We compare \defense against four mitigation baselines.
Filter-based defense~\cite{huiPLeakPromptLeaking2024} removes
leaked prompt content from model outputs.
Sandwich Defense~\cite{learnprompting_sandwich_defense} places
protective instructions before and after user queries.
Prompt Obfuscation~\cite{papePromptObfuscationLarge} optimizes
alternative system-prompt representations to reduce
extractability. 
PromptKeeper~\cite{jiangPromptKeeperSafeguardingSystem2025a}
uses statistical testing to detect leakage and regenerates
flagged responses with a dummy system prompt. These baselines span input reinforcement, prompt protection, output filtering, and response regeneration, covering distinct intervention points and access requirements.

\subsection{Datasets and Workload Construction}
\label{sec:datasets}

\mypara{Prompt sets.}
The evaluation of \defense is conducted using randomly shuffled query streams comprising benign and adversarial prompt behaviors. We construct these streams from three sources: general single-turn benign prompts, multi-turn conversational dialogues, and adversarial prompts designed for prompt leakage attacks.

\mysubpara{Benign prompts.}
Following prior prompt-leakage evaluations~\cite{huiPLeakPromptLeaking2024,effectivepromptextractionlanguage}, we draw benign prompts from GSM8K~\cite{gsm8k}, SST-2~\cite{sst2}, SQuAD v2.0~\cite{SQuAD}, Dolly-15k~\cite{dolly}, and the benign subset of WildJailbreak~\cite{wildteaming2024}. These datasets cover mathematical reasoning, sentiment classification, question answering, and instruction-following tasks, providing diverse semantic patterns representative of real-world LLM usage.

\mysubpara{Multi-turn conversational data.}
To evaluate robustness under realistic dialogue patterns, we use OASST1~\cite{oasst1}, ChatAlpaca~\cite{chatalpaca}, and WildChat~\cite{wildchat}, which contain an average of 8.58, 4.32, and 12.48 turns per conversation, respectively. These datasets contain multi-turn conversations with natural semantic continuity, allowing us to test whether \defense can distinguish benign conversational coherence from adversarial extraction behavior.

\mysubpara{Adversarial prompts.}
Adversarial prompts are drawn from the default prompt sets released with Zhang et al.~\cite{effectivepromptextractionlanguage}, PromptFuzz~\cite{promptfuzz}, LeakAgent~\cite{leakAgent}, and PLeak~\cite{huiPLeakPromptLeaking2024}. We additionally include prompt extraction corpora from WildJailbreak~\cite{wildjailbreak}, AdvPrompter~\cite{advprompter}, Salesforce-Leak~\cite{agarwalPromptLeakageEffect2024}, AI-curated prompts~\cite{GPT}, and human-crafted prompts~\cite{toxicchat,promptfuzz}. These sources introduce broader attack strategies and prompt formulations beyond method-specific datasets. All prompts used in our evaluation are in English.

\mysubpara{Mixed query workload.}
We construct mixed-query workloads by combining a benign query set $\mathcal{B}$ with an adversarial query set $\mathcal{A}$. All adversarial queries are retained to preserve complete attack behavior and are mixed with benign queries to obscure extraction intent. We set $\rho=0.10$, corresponding to a 1:9 adversarial-to-benign ratio. The combined stream is shuffled using a fixed random seed. Results are averaged over multiple benign samples to reduce sampling bias.

\subsection{Evaluation Metrics}
\label{sec:metrics}
\mypara{Detection effectiveness.}
We measure \defense detection performance using standard classification metrics. Specifically, we report: (i) the false positive rate (FPR), which measures the fraction of benign queries incorrectly flagged as attacks; (ii) recall, which measures the fraction of adversarial queries that are successfully detected; and (iii) the F1 score, which captures the trade-off between false positives and false negatives.

\mypara{Prompt leakage success.}
To measure how much sensitive information is extracted, we integrate \defense into the attack frameworks and compare reconstructed outputs against ground-truth system prompts. Following prior work, we use prompts from \emph{awesome-chatgpt-prompts}~\cite{awesome-chatgpt-prompts} as controlled secret prompts and evaluate leakage success for both PLeak~\cite{huiPLeakPromptLeaking2024} and LeakAgent~\cite{leakAgent}. We report four metrics: \emph{substring match} (SM), which flags success when the true prompt appears as a contiguous substring after normalization; \emph{exact match} (EM), which requires identical reconstruction; \emph{semantic similarity} (SS), computed as cosine similarity between sentence embeddings of the true and reconstructed prompts; and \emph{ROUGE}~\cite{rouge}, which measures n-gram overlap.

\mypara{Latency overhead.}
We report per-query latency at p50, p95, and p99 to capture median and tail latency introduced by \defense under online deployment.

\section{Evaluation}
\label{sec:evaluation}

Using the experimental setup in Section~\ref{sec:implementation}, we address four research questions and then analyze the contribution of individual components.

\begin{enumerate}

\item \textbf{RQ1: Detection effectiveness.} Can \defense accurately detect extraction queries in benign-heavy and benign multi-turn workloads while maintaining a low false-positive rate?

\item \textbf{RQ2: Adaptive leakage mitigation.} How much prompt leakage remains when response-adaptive attackers interact with the defended pipeline?

\item \textbf{RQ3: Prompt-only detection.} Can prompt-level signals detect extraction attempts before structural patterns emerge?

\item \textbf{RQ4: Runtime overhead.} What latency and throughput overhead does \defense incur in the evaluated online pipeline?

\end{enumerate}


\subsection{\defense's Detection Effectiveness} \label{sec:rq1}

\begin{table*}[t]
\centering
\scriptsize
\caption{Detection performance of \defense and baseline defenses under mixed-query stream (benign + adversarial).}
\vspace{-4pt}
\label{tab:llama2-AdvGuard-detection-performance}
\resizebox{\textwidth}{!}{
\begin{tabular}{l c | ccc | ccc | ccc}
\toprule
\textbf{Benign Stream} & \textbf{AdvQueries}
& \multicolumn{3}{c}{\textbf{PromptGuard}}
& \multicolumn{3}{c}{\textbf{LLMGuard}}
& \multicolumn{3}{c}{\textbf{\defense}} \\
\cmidrule(lr){3-5}\cmidrule(lr){6-8}\cmidrule(lr){9-11}
&
& \textbf{FPR(\%)$\downarrow$} & \textbf{Recall(\%)$\uparrow$} & \textbf{F1(\%)$\uparrow$}
& \textbf{FPR(\%)$\downarrow$} & \textbf{Recall(\%)$\uparrow$} & \textbf{F1(\%)$\uparrow$}
& \textbf{FPR(\%)$\downarrow$} & \textbf{Recall(\%)$\uparrow$} & \textbf{F1(\%)$\uparrow$} \\
\midrule
\multicolumn{11}{l}{\textbf{Attack: Zhang (COLM'24)}} \\
\midrule
GSM8K & 11234 & 10.75 & 96.90 & 82.91 & 2.37 & 92.75 & 92.73 & \textbf{3.40} & \textbf{99.95} & \textbf{98.11} \\
SST2 & 12346 & 8.85 & 95.75 & 83.22 & 0.68 & 96.40 & 96.93 & \textbf{0.00} & \textbf{100.00} & \textbf{100.00} \\
SQuAD\_v2 & 12824 & 5.73 & 94.50 & 86.88 & 2.12 & 92.25 & 92.09 & \textbf{0.95} & \textbf{92.13} & \textbf{95.88} \\
Dolly15k & 9397 & 3.92 & 96.50 & 94.22 & 4.47 & 92.60 & 91.43 & \textbf{0.00} & \textbf{100.00} & \textbf{100.00} \\
Wildjailbreak & 12055 & 10.10 & 96.30 & 82.75 & 3.21 & 92.00 & 90.60 & \textbf{0.00} & \textbf{91.14} & \textbf{95.12} \\
\midrule
\multicolumn{11}{l}{\textbf{Attack: PromptFuzz}} \\
\midrule
GSM8K & 11234 & 10.56 & 94.66 & 83.24 & 5.81 & 94.52 & 88.82 & \textbf{3.40} & \textbf{99.91} & \textbf{98.23} \\
SST2 & 12346 & 7.40 & 90.99 & 83.25 & 3.29 & 93.85 & 91.41 & \textbf{0.00} & \textbf{99.02} & \textbf{99.51} \\
SQuAD\_v2 & 12824 & 5.00 & 94.28 & 88.19 & 1.46 & 92.80 & 93.63 & \textbf{0.89} & \textbf{94.18} & \textbf{96.92} \\
Dolly15k & 9397 & 11.00 & 93.66 & 86.05 & 6.63 & 93.66 & 90.12 & \textbf{0.00} & \textbf{99.08} & \textbf{99.54} \\
Wildjailbreak & 12055 & 10.69 & 91.47 & 79.63 & 7.01 & 92.76 & 85.16 & \textbf{0.08} & \textbf{93.27} & \textbf{96.08} \\
\midrule
\multicolumn{11}{l}{\textbf{Attack: LeakAgent}} \\
\midrule
GSM8K & 11234 & 4.15 & 35.48 & 50.41 & 2.18 & 71.34 & 81.98 & \textbf{2.75} & \textbf{97.02} & \textbf{97.06} \\
SST2 & 12346 & 7.97 & 34.92 & 47.46 & 5.52 & 72.42 & 80.07 & \textbf{0.00} & \textbf{98.73} & \textbf{99.23} \\
SQuAD\_v2 & 12824 & 11.07 & 35.34 & 46.05 & 3.34 & 71.36 & 80.78 & \textbf{0.66} & \textbf{94.09} & \textbf{97.02} \\
Dolly15k & 9397 & 7.14 & 35.26 & 49.66 & 3.78 & 71.92 & 82.03 & \textbf{0.02} & \textbf{99.51} & \textbf{99.08} \\
Wildjailbreak & 12055 & 11.67 & 35.10 & 46.07 & 2.42 & 71.62 & 81.88 & \textbf{0.17} & \textbf{94.21} & \textbf{96.89} \\
\midrule
\multicolumn{11}{l}{\textbf{Attack: PLeak}} \\
\midrule
GSM8K & 11234 & 4.85 & 57.40 & 70.19 & 4.22 & 78.36 & 85.29 & \textbf{2.75} & \textbf{98.52} & \textbf{96.67} \\
SST2 & 12346 & 3.35 & 56.72 & 70.13 & 4.37 & 78.14 & 84.62 & \textbf{0.00} & \textbf{98.73} & \textbf{99.23} \\
SQuAD\_v2 & 12824 & 3.71 & 56.82 & 69.88 & 4.59 & 77.88 & 84.09 & \textbf{0.25} & \textbf{94.39} & \textbf{94.83} \\
Dolly15k & 9397 & 12.51 & 56.84 & 67.35 & 5.48 & 77.32 & 84.86 & \textbf{0.11} & \textbf{97.69} & \textbf{98.37} \\
Wildjailbreak & 12055 & 6.33 & 57.16 & 68.76 & 4.27 & 78.48 & 84.95 & \textbf{0.06} & \textbf{98.15} & \textbf{99.07}  \\
\bottomrule
\end{tabular}
}
\end{table*}


We address RQ1 by evaluating \defense’s ability to detect prompt-leakage queries under realistic mixed-traffic, multi-turn conversational, and heterogeneous attack workloads.

\paragraph{\textbf{Detection Under Mixed Benign-Adversarial Query Workloads}}

We construct a mixed-query workload by injecting queries from four state-of-the-art prompt-leakage attacks (Zhang \emph{et al.}, PromptFuzz, LeakAgent, and PLeak) into five benign workloads (GSM8K, SST-2, SQuAD\_v2, Dolly15k, and WildJailbreak) following the mixed-query construction policy in Section~\ref{sec:implementation}. This setting models attacks embedded within benign user activity. Table~\ref{tab:llama2-AdvGuard-detection-performance} reports FPR, recall, and F1 score.

\mypara{Results.}
\defense consistently outperforms PromptGuard and LLMGuard in reported F1 across all workloads and attack families, achieving F1 scores of 94.83--100.00\%. \defense FPR remains at or below 3.40\% and is less than 1\% in most settings, while recall ranges from 91.14--100.00\%. The largest performance gap appears under
LeakAgent, where PromptGuard recall drops to approximately \textbf{35\%} and LLMGuard reaches approximately \textbf{71\%}, while \defense maintains 94.09--99.51\%. Performance is also stable across benign workloads, with F1 standard deviation at most 2.04 percentage points for every attack. Appendix~\ref{sec:leave-one-attack-family} evaluates unseen-family transfer.

\mypara{Key Insight.} These results show that \defense reliably detects prompt-leakage attempts hidden within benign traffic. By modeling \textit{interaction-level query structure} rather than isolated prompts, \defense achieves high detection accuracy with low false positives, making it suitable for real-world LLM deployment. Next, we evaluate robustness under realistic interaction patterns. We stress-test \defense under multi-turn conversational streams.

\paragraph{\textbf{Detection Under Multi-Turn Conversational Workloads}}
\label{sec:multiturn}


Real-world LLM interactions are inherently multi-turn and context-dependent. Benign follow-up queries can resemble coordinated probing because both exhibit semantic continuity. To test this setting, we mixed the adversarial queries into three multi-turn conversational datasets: OASST1~\cite{oasst1}, ChatAlpaca~\cite{chatalpaca}, and WildChat~\cite{wildchat}.

\begin{table}[t]
\centering
\small
\setlength{\tabcolsep}{4pt}
\renewcommand{\arraystretch}{0.85}
\caption{Detection performance under multi-turn conversational streams. Metrics are reported in \%}
\label{tab:multiturn_results}
\vspace{-4pt}
\resizebox{\columnwidth}{!}{
\begin{tabular}{l c | ccc | ccc | ccc}
\toprule
\textbf{Dataset} & \textbf{Queries}
& \multicolumn{3}{c}{\textbf{PromptGuard}}
& \multicolumn{3}{c}{\textbf{LLMGuard}}
& \multicolumn{3}{c}{\textbf{\defense}} \\
\cmidrule(lr){3-5} \cmidrule(lr){6-8} \cmidrule(lr){9-11}
&
& \textbf{FPR} & \textbf{Recall} & \textbf{F1}
& \textbf{FPR} & \textbf{Recall} & \textbf{F1}
& \textbf{FPR} & \textbf{Recall} & \textbf{F1} \\
\midrule
OASST1    
& 8246 
& 4.74 & 86.45 & 91.92 
& 1.47 & 89.85 & 94.41 
& \textbf{1.34} & \textbf{100.00} & \textbf{99.12} \\

ChatAlpaca  
& 9276 
& 3.77 & 87.68 & 92.41 
& 3.98 & 89.52 & 93.38 
& \textbf{1.76} & \textbf{100.00} & \textbf{98.12} \\

WildChat 
& 8054 
& 10.67 & 85.92 & 90.64 
& 3.63 & 87.75 & 92.88 
& \textbf{3.42} & \textbf{100.00} & \textbf{97.23} \\

\bottomrule
\end{tabular}
}
\end{table}

\mypara{Results.}
As shown in Table~\ref{tab:multiturn_results}, \defense maintains strong detection performance across all conversational workloads. \defense consistently outperforms PromptGuard and LLMGuard across all conversational workloads, achieving 97.23--99.12\% F1 and \textbf{100\% recall}. FPR remains low at 1.34--3.42\%, compared with up to 10.67\% for PromptGuard, while both baselines achieve only 85--90\% recall. Appendix~\ref{sec:hetro-attack-benign} evaluates mixed attack families.




\mypara{Key Insight.}
Overall, \defense effectively distinguishes benign conversational coherence from adversarial extraction patterns, maintaining robust detection in multi-turn settings.

\subsection{Adaptive leakage mitigation.}
\label{sec:rq2}

\begin{table*}[t]
\centering
\small
\setlength{\tabcolsep}{4.2pt}
\renewcommand{\arraystretch}{1.08}
\caption{Prompt leakage scores for PLEAK and LeakAgent adaptive querying attacks across victim models and defenses. Lower values indicate stronger leakage mitigation.}
\label{tab:leakage-defense}
\vspace{-4pt}
\resizebox{\textwidth}{!}{%
\begin{tabular}{l l cccc cccc cccc cccc}
\toprule
\textbf{Attack} & \textbf{Defense}
& \multicolumn{4}{c}{\textbf{GPT-2}}
& \multicolumn{4}{c}{\textbf{Falcon3}}
& \multicolumn{4}{c}{\textbf{LLaMA3.2}}
& \multicolumn{4}{c}{\textbf{Qwen2.5}} \\
\cmidrule(lr){3-6} \cmidrule(lr){7-10} \cmidrule(lr){11-14} \cmidrule(lr){15-18}
& &
\textbf{SM}$\downarrow$ & \textbf{EM}$\downarrow$ & \textbf{SS}$\downarrow$ & \textbf{ROUGE-L}$\downarrow$
& \textbf{SM}$\downarrow$ & \textbf{EM}$\downarrow$ & \textbf{SS}$\downarrow$ & \textbf{ROUGE-L}$\downarrow$
& \textbf{SM}$\downarrow$ & \textbf{EM}$\downarrow$ & \textbf{SS}$\downarrow$ & \textbf{ROUGE-L}$\downarrow$
& \textbf{SM}$\downarrow$ & \textbf{EM}$\downarrow$ & \textbf{SS}$\downarrow$ & \textbf{ROUGE-L}$\downarrow$ \\
\midrule

\multirow{3}{*}{\textbf{PLEAK}}
& No Defense 
& 0.74 & 0.73 & 0.92 & 0.95
& 0.78 & 0.73 & 0.81 & 0.91
& 0.92 & 0.92 & 0.99 & 0.98 
& 0.97 &  0.95 & 0.99 & 0.97\\
& Sandwich-based
& 0.48 & 0.28 & 0.82 & 0.71
& 0.64 & 0.55 & 0.81 & 0.83
& 0.61 & 0.52 & 0.82 & 0.71
& 0.91 & 0.90 & 0.94 & 0.92 \\
& Filter-based
& 0.09 & 0.01 & 0.41 & 0.47
& 0.20 & 0.04 & 0.53 & 0.64
& 0.14 & 0.03 & 0.48 & 0.54
& 0.29 & 0.11 & 0.61 & 0.66 \\
& Prompt Obfuscation 
& 0.01 & 0.01 & 0.09 & 0.13
& 0.01 & 0.01 & 0.09 & 0.19
& 0.02 & 0.01 & 0.17 & 0.31
& 0.01 & 0.00 & 0.22 & 0.29 \\
& PromptKeeper 
& 0.00 & 0.00 & 0.04 & 0.12
& 0.00 & 0.00 & 0.07 & 0.18
& 0.01 & 0.00 & 0.14 & 0.28
& 0.01 & 0.00 & 0.22 & 0.26 \\
& \textbf{\defense}
& 0.01 & 0.00 & 0.05 & 0.12
& 0.01 & 0.00 & 0.12 & 0.30
& 0.02 & 0.00 & 0.16 & 0.21
& 0.02 & 0.01 & 0.21 & 0.23 \\
\midrule

\multirow{3}{*}{\textbf{LeakAgent}}
& No Defense
& 0.90 & 0.89 & 0.90 & 0.93
& 0.85 & 0.84 & 0.95 & 0.95
& 0.94 & 0.89 & 0.97 & 0.95
& 0.98 & 0.95 & 0.99 & 0.94 \\
& Sandwich-based
& 0.37 & 0.22 & 0.64 & 0.68
& 0.72 & 0.63 & 0.75 & 0.91
& 0.81 & 0.75 & 0.85 & 0.89
& 0.95 & 0.93 & 0.99 & 0.91 \\
& Filter-based
& 0.09 & 0.00 & 0.33 & 0.34
& 0.16 & 0.11 & 0.56 & 0.60
& 0.12 & 0.09 & 0.38 & 0.44
& 0.33 & 0.19 & 0.63 & 0.66 \\
& Prompt Obfuscation 
& 0.01 & 0.01 & 0.17 & 0.35
& 0.02 & 0.02 & 0.26 & 0.40
& 0.03 & 0.01 & 0.14 & 0.25
& 0.14 & 0.04 & 0.30 & 0.61 \\
& PromptKeeper 
& 0.00 & 0.00 & 0.01 & 0.14
& 0.00 & 0.00 & 0.06 & 0.19
& 0.02 & 0.01 & 0.22 & 0.37
& 0.03 & 0.00 & 0.21 & 0.34 \\
& \textbf{\defense}
& 0.01 & 0.00 & 0.02 & 0.14
& 0.01 & 0.00 & 0.15 & 0.15
& 0.01 & 0.00 & 0.16 & 0.17
& 0.09 & 0.01 & 0.25 & 0.52 \\
\bottomrule
\end{tabular}%
}
\end{table*}

Building on the detection performance results in Section~\ref{sec:rq1}, we evaluate whether \defense can prevent system prompt reconstruction when deployed directly within the online inference pipeline, addressing \textbf{RQ2}.


We integrate \defense into two adaptive prompt-leakage attacks, \textbf{PLEAK} and \textbf{LeakAgent}, which refine subsequent queries based on prior interactions. Suspicious queries receive a standardized denial response (``Request denied due to safety policy''), limiting further prompt exposure. This setting evaluates response-adaptive attacks rather than static attack templates.

We compare \defense against filter-based defense~\cite{filtering_defense}, Sandwich Defense~\cite{sandwich_defense}, Prompt Obfuscation~\cite{papePromptObfuscationLarge}, and PromptKeeper~\cite{jiangPromptKeeperSafeguardingSystem2025a}. Unlike defenses requiring privileged information, \defense uses incoming queries without access to the system prompt, model internals, or generated responses. Table~\ref{tab:leakage-defense} reports leakage across GPT-2~\cite{radford2019language}, Falcon3~\cite{falcon3}, LLaMA~3.2~\cite{llama3_8b}, and Qwen2.5~\cite{qwen2.5}.

\mypara{Results.}
Table~\ref{tab:leakage-defense} shows that \defense limits exact match (EM) to 0.00--0.01 and substring match (SM) to 0.01--0.09 across both adaptive attacks
and all four victim models, compared with 0.73--0.95 EM and 0.74--0.98 SM without defense. Semantic similarity remains at 0.02--0.25, and both
semantic similarity and ROUGE-L are lower than those of sandwich- and filter-based defenses in every setting. Despite operating solely on incoming user queries, \defense also surpasses PromptKeeper on ROUGE-L under LeakAgent on Falcon3 (0.15 vs.\ 0.19) and LLaMA 3.2 (0.17 vs.\ 0.37), although PromptKeeper retains an advantage on Qwen 2.5 (0.34 vs.\ 0.52). These results demonstrate substantial leakage mitigation without access to system prompts, model internals, or generated responses.

\mypara{Key Insight.} 
\defense effectively limits adaptive prompt reconstruction by detecting the progressive query patterns of extraction attacks. The principal advantage of \defense is its combination of limited access requirements and consistently lower leakage.


\subsection{Detection of Prompt Leakage in Sparse Adversarial Regimes}
\label{sec:rq3}


Prompt-leakage attacks may begin with a small number of engineered probes before sufficient interaction history emerges for structural analysis~\cite{alizadeh2025simplepromptinjectionattacks, Prompt_Engineering_Guide_2022}. \textbf{RQ3} evaluates whether \defense can detect such early-stage attempts using only prompt-level signals. We disable graph-based analysis and evaluate each query independently.


\begin{table}[t]
\centering
\caption{Early-stage prompt-level detection performance of \textbf{\defense} when graph-based structural analysis is disabled.}
\label{tab:early-detection-for-prompt-leakage}
\vspace{-4pt}
\footnotesize
\setlength{\tabcolsep}{2pt}
\begin{tabular}{l l r r r r}
\toprule
\textbf{Benign} & \textbf{Adversarial} & \textbf{\#AdvQry} & \textbf{FPR(\%)$\downarrow$} & \textbf{Recall(\%)$\uparrow$} & \textbf{F1(\%)$\uparrow$} \\
\midrule
GSM8K          & WildJailbreak   & 543  & 0.08 & 96.93 & 97.99 \\
SST-2           & AdvPrompter     & 2004  & 1.20 & 95.00 & 97.00 \\
SQuAD\_v2      & SalesforceLeak & 841 & 0.09 & 96.40 & 97.94 \\
Dolly15k       & AI-curated       & 958 & 0.00 & 99.67 & 99.35 \\
JailbreakBench & humanAI-Craft     & 482  & 0.00 & 99.64 & 99.35 \\
\bottomrule
\end{tabular}
\end{table}



We construct five settings by combining benign~\cite{gsm8k,sst2,SQuAD,dolly,wildjailbreak} and adversarial query sets~\cite{wildjailbreak, advprompter, toxicchat, wildteaming2024, agarwalPromptLeakageEffect2024} drawn from the sources described in Section~\ref{sec:implementation}, at a 1:9 adversarial-to-benign ratio. These include benchmark attacks and human- and model-crafted prompts targeting system instructions (e.g., instruction override and prompt disclosure queries). We classify queries as adversarial when their prompt-level risk score exceeds a threshold \( \theta_p = 0.8 \) while benign queries typically fall within the range $[0.05, 0.40]$.

\mypara{Results.} 
Table~\ref{tab:early-detection-for-prompt-leakage} shows that \defense achieves a high recall of over \textbf{95\%} and F1 scores up to \textbf{99.35\%} across all settings. In the meantime, \defense maintains low false-positive rates at no more than \textbf{1.20\%}. Performance remains stable across diverse adversarial prompt sources, including both benchmark datasets and curated attack prompts.

\mypara{Key Insight.} This demonstrates that \defense can detect early-stage prompt-leakage attempts, preventing attackers from conducting slow and stealthy prompt-leakage attempts before larger structural patterns emerge.


\subsection{Runtime overhead}
\label{sec:rq4}


\textbf{RQ4} evaluates the runtime overhead of \defense in online LLM inference. We compare \defense with \textsc{PromptGuard} and \textsc{LLMGuard} on LLaMA3.2 with sequence length 512 over 1,000 consecutive queries, reporting average and tail latency (P50, P95, P99) and throughput (queries/s).

\begin{figure}[t]
  \centering
  \includegraphics[width=0.98\columnwidth]{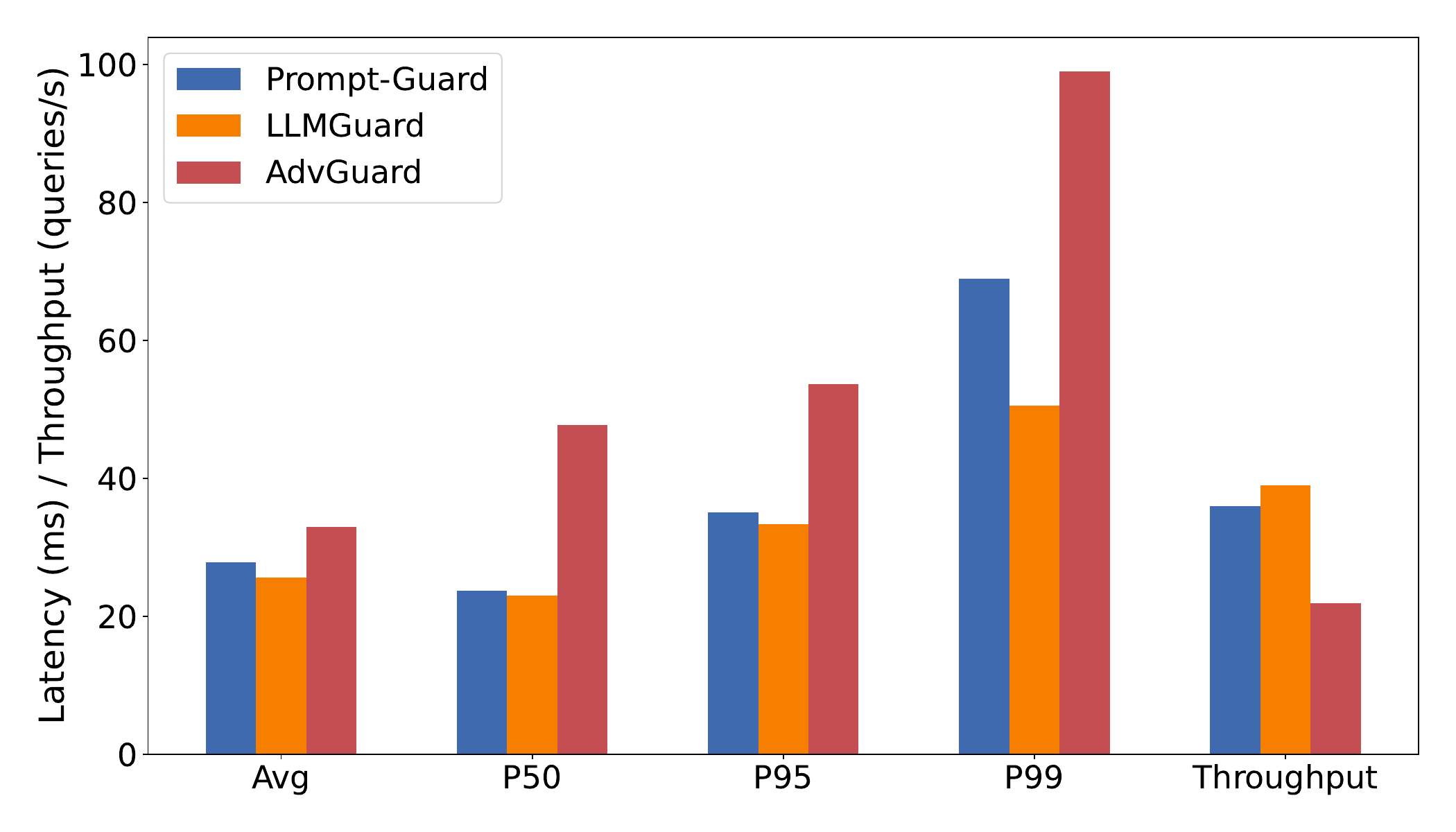}
  \vspace{-10pt}
  \caption{Latency and throughput comparison between \defense, \textsc{PromptGuard}, and \textsc{LLMGuard}.}
  \label{fig:latency_profile-fig}
  \vspace{-8pt}
\end{figure}

\begin{figure}[t]
  \centering
  \includegraphics[width=0.98 \columnwidth]{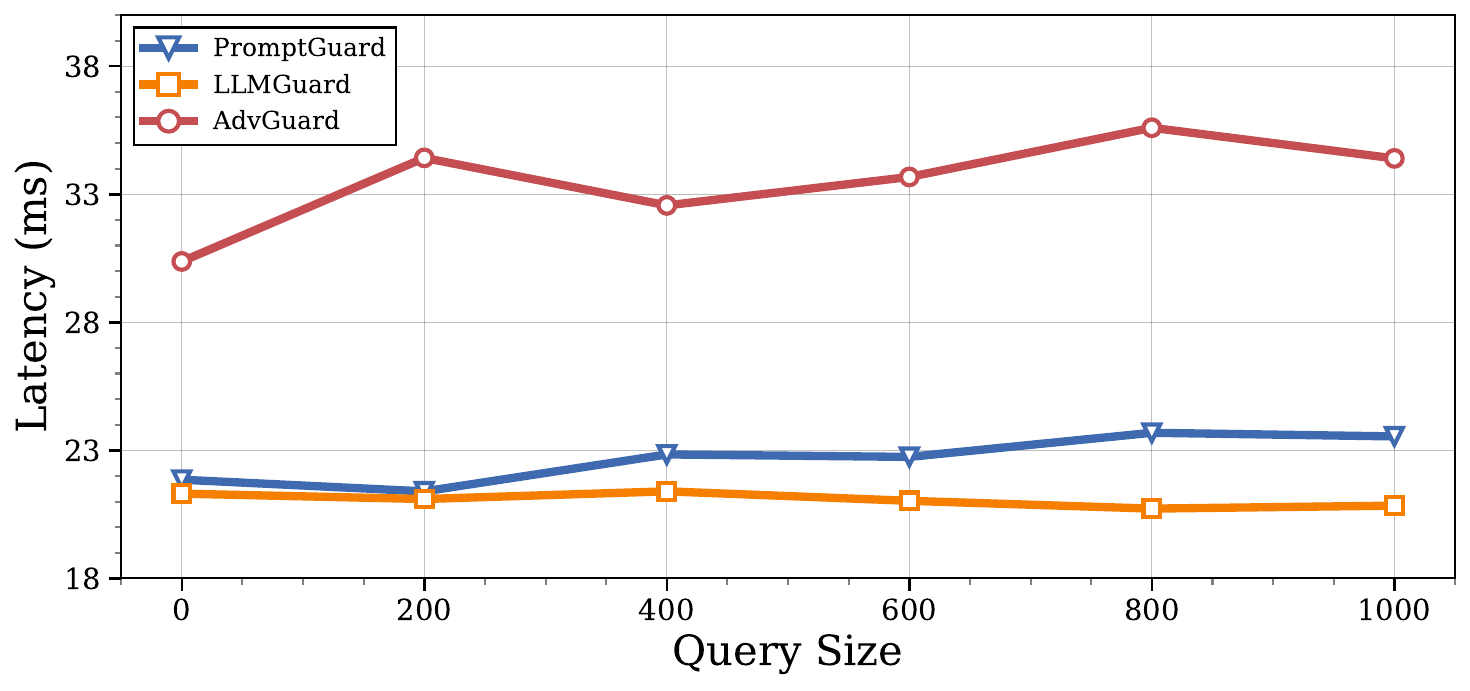}
  \vspace{-10pt}
  \caption{Latency stability of \defense compared with \textsc{PromptGuard} and \textsc{LLMGuard}. Each point represents the average latency of 200 queries.}
  \label{fig:latency_for-query-seq}
  \vspace{-8pt}
\end{figure}

\mypara{Overall performance.}
As shown in Figure~\ref{fig:latency_profile-fig}, \defense introduces moderate overhead compared with stateless baselines. Average latency is \textbf{32.96 ms/query}, with \textbf{54.7 ms P95} and \textbf{98.9 ms P99} latency, while throughput decreases from 35--39 to \textbf{21.9 queries/s}. This overhead primarily stems from interaction-aware processing, including structural analysis. A detailed component-level latency breakdown is provided in Appendix~\ref{sec:runtime_breakdown}. Despite this overhead, the absolute latency remains practical for real-time deployment.

\mypara{Stability under long sessions.}
Despite its stateful architecture, Figure~\ref{fig:latency_for-query-seq} shows that \defense maintains stable latency across 1,000 queries, demonstrating bounded runtime overhead during long-running sessions. This stability is achieved through bounded graph state management, selective caching, and eviction policies, which collectively ensure predictable runtime behavior during long-running sessions.

\mypara{Key Insight.} While stateless defenses achieve lower latency, they provide substantially weaker protection against adaptive and coordinated attacks (as demonstrated in Sections~\ref{sec:rq1}--\ref{sec:rq2}). \defense intentionally trades a modest, well-bounded runtime overhead for significantly stronger security guarantees, making it a practical and deployable defense for production-scale LLM systems.

\subsection{Ablation Study}
\label{sec:ablation}

We ablate \defense to assess three hypotheses: (i) interaction-level modeling is necessary beyond per-query features, (ii) no detection stage is sufficient in isolation, and (iii) structural learning is required to capture coordinated extraction behavior. We systematically remove or simplify pipeline components and measure their impact on detection performance.

\begin{figure}[t]
    \centering
    \begin{minipage}[t]{0.48\columnwidth}
        \centering
        \includegraphics[width=\linewidth]{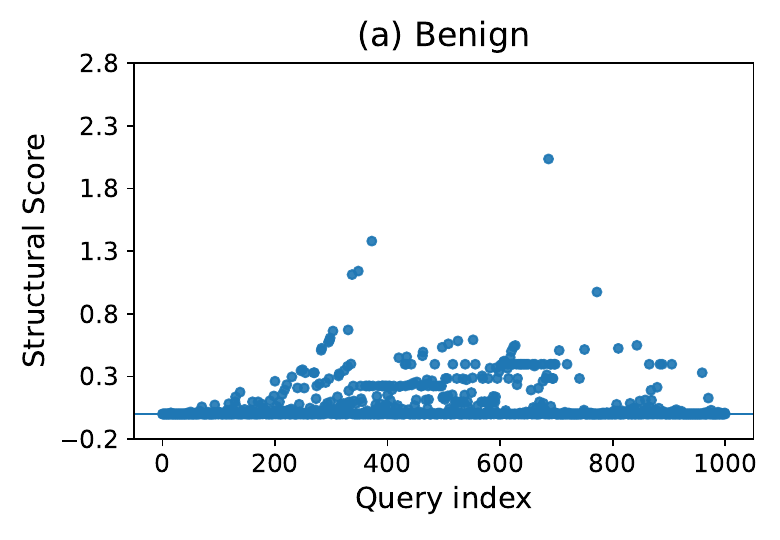}
    \end{minipage}
    \hfill
    \begin{minipage}[t]{0.48\columnwidth}
        \centering
        \includegraphics[width=\linewidth]{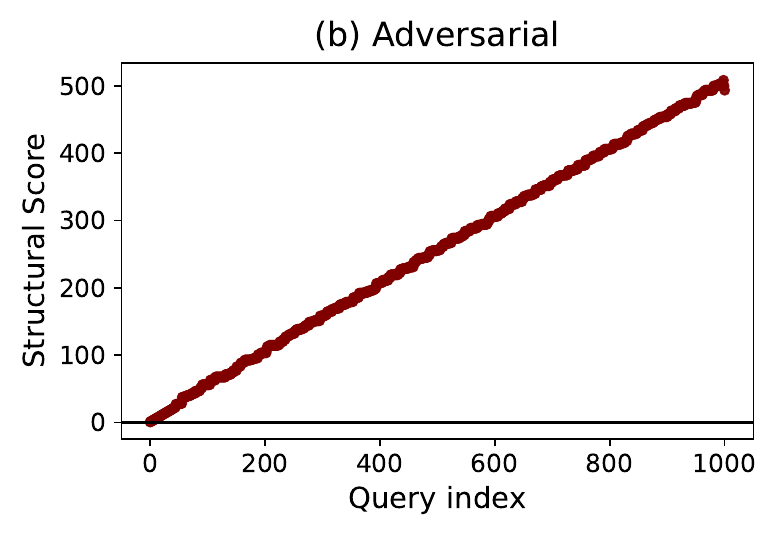}
    \end{minipage}
    \vspace{-10pt}
    \caption{Weighted sum distribution for benign and adversarial queries over time.}
    \label{fig:structural-risk-score}
\end{figure}

\mysubpara{Sequence-level modeling necessary for detecting prompt leakage?}
We test whether reliable prompt-leakage detection requires modeling interaction structure beyond individual queries.
Figure~\ref{fig:structural-risk-score} shows aggregated similarity scores over a mixed-query stream. Benign scores remain unstructured, with 88\% of $s_C$ values below 0.5 and no temporal trend. Adversarial scores increase monotonically as semantic reuse accumulates across turns. Such signals are not observable at the level of individual queries $q_t$ — each appears benign in isolation, but separation emerges only when similarities are aggregated over the interaction history $\{q_1,\ldots,q_t\}$.

\begin{figure}[t]
  \centering
  \vspace{-10pt}
  \includegraphics[width=0.70\columnwidth]{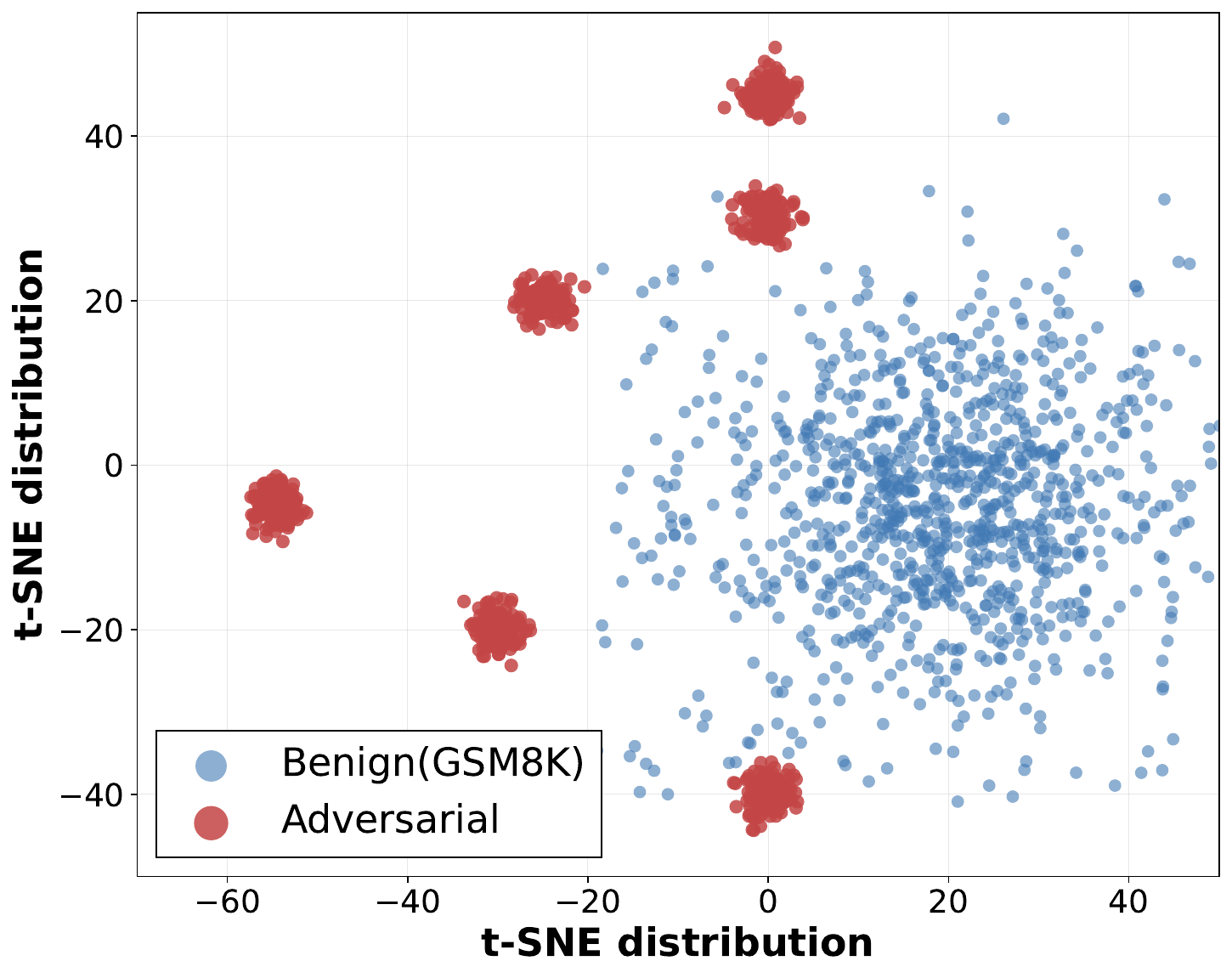}
  \vspace{-10pt}
  \caption{t-SNE projection of benign and adversarial query embeddings.}
  \label{fig:tsne_visualization}
\end{figure}

To further analyze separability, we project sequence-level graph embeddings using t-SNE (Figure~\ref{fig:tsne_visualization}). Benign interactions form diffuse, overlapping clusters, whereas adversarial sequences concentrate into tight, well-separated regions. This separation does not arise from individual query embeddings, but from the structural relationships encoded in the interaction graph, indicating that sequence-level representations provide significantly stronger discriminative power.


\mypara{Key Insight.}
Prompt leakage is a structural phenomenon: adversarial intent spans semantically related queries, making per-query detection insufficient. \defense's interaction graph $G_t=(V_t,E_t)$ reveals this coordination, motivating the stronger structural reasoning analyzed next.

\mysubpara{Is a Multi-Stage Detection Pipeline Necessary?} 
Table~\ref{tab:ablation_components} evaluates the contribution of each detection stage in \defense.

\begin{table}[t]
\centering
\small
\caption{Component-wise ablation of \defense. $\Delta$ denotes the relative change from the prompt-only baseline.}
\label{tab:ablation_components}
\vspace{-4pt}
\resizebox{\columnwidth}{!}{%
\begin{tabular}{l l
    r @{\hspace{4pt}} r
    r @{\hspace{4pt}} r
    r @{\hspace{4pt}} r}
\toprule
\textbf{Dataset}
  & \textbf{Method}
  & \textbf{FPR(\%)}$\downarrow$ & $\Delta$
  & \textbf{Recall(\%)}$\uparrow$ & $\Delta$
  & \textbf{F1(\%)}$\uparrow$ & $\Delta$ \\
\midrule

\multirow{3}{*}{OASST1}
  & Prompt Signal     & 11.50 & ---     & 69.00  & ---     & 77.00  & --- \\

  & Outlier Filter    & \textbf{0.00} & -11.50 & 100.00 & +31.00 & 100.00 & +23.00 \\
  & Graph Classifier  & \textbf{0.00} & -11.50 & 100.00 & +31.00 & 100.00 & +23.00 \\
\midrule

\multirow{3}{*}{WildChat}
  & Prompt Signal     & 18.30 & ---     & 69.00  & ---     & 74.00  & --- \\

  & Outlier Filter    & 6.70  & -11.60 & 100.00 & +31.00 & 97.00  & +23.00 \\
  & Graph Classifier  & \textbf{3.18} & -15.12 & 100.00 & +31.00 & \textbf{98.00} & +24.00 \\
\midrule

\multirow{3}{*}{ChatAlpaca}
  & Prompt Signal     & 1.88 & ---     & 69.00 & ---     & 81.00 & --- \\

  & Outlier Filter    & 3.46 & +1.58  & 100.00 & +31.00 & 98.00  & +17.00 \\
  & Graph Classifier  & \textbf{0.00} & -1.88 & 100.00 & +31.00 & \textbf{100.00} & +19.00 \\
\bottomrule
\end{tabular}%
}
\end{table}

\paragraph{Prompt-level detection is insufficient}
Using only the prompt signal (Stage 1) yields moderate F1 scores (74--81\%) but consistently low recall (69\%) across all datasets, indicating limited detection of attacks distributed across queries.


\paragraph{Statistical filtering reduces false positives but is insufficient alone}
Adding structural aggregation raises recall to \textbf{100\%} and F1 from \textbf{74--81\%} to \textbf{97--100\%}, showing the benefit of interaction-level modeling. However, false positives remain higher on conversational workloads such as WildChat due to ambiguity between adversarial sequences and semantically coherent benign interactions.


\paragraph{Graph-based reasoning reduces residual structural ambiguity}
The graph classifier further improves robustness, preserving \textbf{100\%} recall while reducing or matching the filter's FPR across datasets. On WildChat, for example, FPR decreases from \textbf{6.70\%} to \textbf{3.18\%}, while F1 improves from \textbf{97\%} to \textbf{98\%}. These results suggest that graph-based reasoning effectively distinguishes coordinated adversarial interaction patterns from benign but topically coherent conversations.

\mypara{Key Insight.}
These results demonstrate that no single stage is sufficient: prompt-level analysis misses distributed attacks, while structural aggregation remains susceptible to benign conversational overlap. Combining prompt, interaction-level, and graph-based signals enables \defense to achieve high recall with low false positives.


\mysubpara{Is Structural Learning Necessary Beyond Scalar Aggregation.}
We test whether scalar aggregation of similarity features suffices for adversarial classification or whether graph-based relational modeling is needed.
We construct interaction graphs from four adversarial datasets by incrementally inserting queries connected via semantic similarity, and compare three backends: \textbf{SimMLP} (scalar features, no connectivity), \textbf{1-GCN}, and \textbf{3-GCN}, trained for 20 epochs on a 70:20:10 split.
Table~\ref{tab:gnn-ablation} shows that SimMLP fails substantially across all datasets, achieving only 40.48\% average accuracy, compared with 93.85\% for 1-GCN and nearly 100\% for 3-GCN. These results support multi-hop relational modeling for capturing coordination that scalar aggregation misses.

\begin{table}[t]
\centering
\caption{Impact of GNN depth on adversarial query detection accuracy over 20 training epochs. All values are reported in percentages.}
\label{tab:gnn-ablation}
\vspace{-4pt}
\resizebox{0.89\columnwidth}{!}{%
\begin{tabular}{@{}lcccc@{}}
\toprule
\textbf{Dataset} & \textbf{Size} & \textbf{SimMLP(\%)} & \textbf{1-GCN(\%)} & \textbf{3-GCN(\%)} \\
\midrule
WildJailbreak    & 100k   & 30.13 & 94.05 & \textbf{99.99} \\
AdvPrompter      & 15,119 & 42.61 & 96.09 & \textbf{100.00} \\
Salesforce-Leak  & 7,780  & 43.25 & 90.34 & \textbf{100.00} \\
JailbreakBench   & 5,855  & 45.94 & 94.90 & \textbf{100.00} \\
\midrule
\textbf{Average} & --     & 40.48 & 93.85 & \textbf{100.00} \\
\bottomrule
\end{tabular}%
}
\end{table}

\mypara{Key Insight.}
Prompt-leakage structure is not linearly separable at the scalar level: relational graph modeling is necessary, not optional, for detecting coordinated multi-turn extraction.

\section{Discussion}\label{sec:discussion}

\defense's design leverages historical interaction patterns to achieve strong detection performance under adaptive, multi-turn prompt-leakage attacks. 

\mypara{Limitations.}
Because \defense relies on interaction structure, meaningful graph structure requires sufficient query history. Very short sessions or extremely sparse attack patterns may therefore evade early detection. The system also depends on semantic embeddings to establish query similarity, which introduces an inductive bias that can reduce sensitivity to highly obfuscated, stylistically diverse queries. Finally, the system requires additional storage for graph structures.   

\mypara{Evasion considerations.}
An adaptive adversary might attempt to circumvent detection by distributing queries across multiple identities or systematically paraphrasing them. However, such strategies inherently inflate the adversary's query budget and reduce the efficiency of information aggregation. Because successful prompt leakage typically requires iterative refinement and cross-query synthesis, eliminating all detectable structural patterns while still making meaningful progress remains difficult in practice.

\mypara{Generalization.}
While our evaluation focuses on prompt leakage, the core principle of detecting malicious behavior via structured query provenance may naturally extend to a broader class of black-box attacks, including model extraction, jailbreak discovery, and data-exfiltration attempts that rely on iterative querying, Appendix~\ref{sec:other_adversarial-attack}. Because \defense is model-agnostic and requires no internal model access, it can be deployed across diverse LLM architectures and updated systems without retraining.

\mypara{Future work.}
Several directions remain promising for improving robustness and scalability. These include exploring hybrid features that combine semantic embeddings with lightweight syntactic or statistical signals to better handle stylistically diverse attacks, investigating longer-horizon or session-linked memory mechanisms, evaluating performance against emerging multi-modal and agentic attack patterns~\cite{sternak}, and studying federated or privacy-preserving provenance aggregation across distributed deployments~\cite{zhuangProxyPromptSecuringSystem2025}.
Overall, \defense does not claim to eliminate prompt leakage. Rather, it shifts the defense boundary from brittle prompt filtering to the significantly harder-to-evade dynamics of interaction structure.

\section{Conclusion}\label{sec:conclusion}

This work demonstrates system prompt leakage as a progressive interaction-level threat rather than a single-query anomaly. Unlike existing defenses that analyze prompts in isolation, \defense detects coordinated extraction behavior through a multi-stage pipeline combining prompt-level risk estimation, interaction-graph construction, statistical filtering, and graph-based classification.

Our evaluation shows that this design improves both detection and mitigation. \defense achieves high recall with low false positives under mixed-query and multi-turn workloads, reduced leakage under adaptive attacks, effective early-stage detection, and bounded runtime overhead. Ablations further show that combining prompt-level, structural, and graph-based signals is necessary for robust detection.

Overall, \defense establishes interaction-level monitoring as a practical and more robust alternative to prompt-only defenses, making system prompt extraction harder by detecting the coordinated query patterns that successful leakage requires.



\IEEEpeerreviewmaketitle

\bibliographystyle{IEEEtran}
\bibliography{references/references}

\appendices

\section{Additional Robustness Evaluations}
\subsection{Leave-One-Attack-Family-Out Evaluation}
\label{sec:leave-one-attack-family}

We evaluated cross-family generalization using a leave-one-attack-family-out protocol. In each fold, one attack family was excluded from training and validation and reserved exclusively for testing. Benign prompts were split into disjoint training, validation, and test sets using a 70/20/10 split, while the remaining attack families were divided into 80\% training and 20\% validation sets. Each fold contained approximately 8,000 training graphs, 1,000 validation graphs, and 2,000 balanced test graphs. On unseen families, Table~\ref{tab:leave-one-family-out} shows \defense achieves 99.98\% mean recall and 99.99\% mean F1 with no false positives.

\begin{table}[!htbp]
\centering
\caption{Leave-one-attack-family-out test results.}
\label{tab:leave-one-family-out}
\small
\vspace{-4pt}
\setlength{\tabcolsep}{5pt}
\begin{tabular}{lccc}
\toprule
\textbf{Held-out family} & \textbf{FPR(\%)$\downarrow$} & \textbf{Rec(\%)$\uparrow$} & \textbf{F1(\%)$\uparrow$} \\
\midrule
LeakAgent    & 0.00 & 100.00 & 100.00 \\
PLeak        & 0.00 &  99.90 &  99.95 \\
PromptFuzz   & 0.00 & 100.00 & 100.00 \\
Zhang et al. & 0.00 & 100.00 & 100.00 \\
\midrule
\textbf{Mean}         & \textbf{0.00} &  \textbf{99.98} &  \textbf{99.99} \\
\bottomrule
\end{tabular}
\end{table}

\subsection{Detection Under Heterogeneous Attacks Mixed With Benign Multi-Turn Conversations}
\label{sec:hetro-attack-benign}

\begin{table}[!htbp]
\centering
\caption{Detection performance of \defense under heterogeneous attack families randomly mixed with benign multi-turn conversations.}
\label{tab:mixed_attack}
\scriptsize
\vspace{-4pt}
\begin{tabular}{l c c c c}
\toprule
\textbf{Benign Stream} & \textbf{Queries} & \textbf{FPR(\%)$\downarrow$} & \textbf{Recall(\%)$\uparrow$} & \textbf{F1(\%)$\uparrow$} \\
\midrule
ChatAlpaca & 9276 & 1.19 & 99.00 & 94.42 \\
OASST1     & 8152 & 0.10 & 99.11 & 99.55 \\
WildChat   & 8056 & 3.40 & 98.00 & 85.74 \\
\bottomrule
\end{tabular}
\end{table}

We evaluate \defense under heterogeneous attacks mixed with benign multi-turn conversations. We combine multiple prompt-leakage attacks~\cite{leakAgent,huiPLeakPromptLeaking2024,promptfuzz,zhangEffectivePromptExtraction2024}, each contributing 25\% of adversarial queries, and randomly mix them with a benign multi-turn conversational dataset drawn from ChatAlpaca~\cite{chatalpaca}, OASST1~\cite{oasst1}, and WildChat~\cite{wildchat}.

Table~\ref{tab:mixed_attack} reports recall of 98.00--99.11\% across the three streams. FPR is 1.19\% on ChatAlpaca, 0.10\% on OASST1, and 3.40\% on WildChat. Despite high recall, WildChat has a lower F1 score of 85.74\%, indicating greater false-positive sensitivity on the more diverse stream. Overall, \defense remains robust under mixed heterogeneous attack families, maintaining high detection accuracy.


\subsection{PLeak Leakage Analysis Across Its Evaluation Datasets}
\label{sec:pleak-leak-analysis}

We evaluate PLeak extraction across domains and system prompts using its original setup~\cite{huiPLeakPromptLeaking2024}, with LLaMA2~\cite{llama2} generating adversarial queries, where PLeak achieves its highest success rates. Due to PLeak's roughly two-hour runtime per dataset, we restrict this evaluation to LLaMA2~\cite{llama2} only. We report the same four leakage metrics used in PLeak—Substring Match (SM), Exact Match (EM), Semantic Similarity (SS), and ROUGE-L scores.

Table~\ref{tab:pleak-defense-results-with-AdvGuard} shows \defense consistently maintains low leakage. SM and EM remain at \textbf{0.00--0.02}, while SS and ROUGE-L range from \textbf{0.08--0.33} across datasets.

\begin{table}[!htbp]
\centering
\small
\setlength{\tabcolsep}{4pt}
\renewcommand{\arraystretch}{0.75}
\caption{PLEAK's performance with \textbf{AdvGuard} using default target system prompts across its evaluation datasets}
\label{tab:pleak-defense-results-with-AdvGuard}
\vspace{-4pt}
\scalebox{0.95}{%
\begin{tabular}{lcccc}
\toprule
\textbf{Dataset} & \textbf{SM}$\downarrow$ & \textbf{EM}$\downarrow$ & \textbf{SS}$\downarrow$ & \textbf{ROUGE-L}$\downarrow$ \\
\midrule
Roles           & 0.00 & 0.00 & 0.12 & 0.14 \\
SIQA            & 0.01 & 0.00 & 0.33 & 0.27 \\
Financial       & 0.02 & 0.02 & 0.15 & 0.11 \\
SQuAD2          & 0.00 & 0.00 & 0.09 & 0.15 \\
GSM8K           & 0.01 & 0.00 & 0.12 & 0.16 \\
ChatGPT-Roles   & 0.02 & 0.00 & 0.15 & 0.17 \\
Rotten Tomatoes & 0.00 & 0.00 & 0.08 & 0.13 \\
\bottomrule
\end{tabular}
}
\end{table}

\section{Additional Design Studies}
\subsection{Nearest-Neighbor vs. Top-$k$ Connectivity}
\label{sec:feature_design_choice}

\defense relies on a similarity graph to capture structural relationships among queries over time. A key design question is how to construct this graph: whether each query should connect to only its closest neighbor, or to multiple similar queries (top-$k$). 
We compared nearest-neighbor connectivity with top-$k$ ($k=3$) graph construction (Table~\ref{tab:graph_construction_ablation}). Top-$k$ raises GSM8K FPR from 3.40\% to 16.35\% and Dolly-15k false negatives from 10 to 462, reducing recall from 99.52\% to 78.00\%. Additional similarity computations and graph updates also increase latency by approximately $3\times$. We therefore adopt second nearest-neighbor connectivity for its stronger overall detection--latency trade-off.

\begin{table}[!htbp]
\centering
\small
\caption{\textbf{Design study for similarity graph connectivity (NN vs. top-$k$). Latency (Lat) is reported in milliseconds.}}
\setlength{\tabcolsep}{4pt}
\vspace{-4pt}
\begin{tabular}{lccccc}
\toprule
\textbf{Data} & \textbf{Method} & \textbf{FPR(\%)$\downarrow$} & \textbf{Rec(\%)$\uparrow$} & \textbf{F1(\%)$\uparrow$} & \textbf{Lat$\downarrow$} \\
\midrule
\multirow{2}{*}{GSM8K}
& Top-$k$     & 16.35 & 100.00 & 93.00 & 92.48 \\
& NN          & \textbf{3.40} & 99.24 & \textbf{98.00} & \textbf{30.55} \\
\midrule
\multirow{2}{*}{SST-2}
& Top-$k$     & 0.00 & 99.48 & 100.00 & 105.48 \\
& NN          & 0.00 & 99.48 & 100.00 & \textbf{32.96} \\
\midrule
\multirow{2}{*}{Dolly}
& Top-$k$     & 0.00 & 78.00 & 88.00 & 86.48 \\
& NN          & 0.00 & \textbf{99.52} & \textbf{100.00} & \textbf{28.45} \\
\bottomrule
\end{tabular}
\label{tab:graph_construction_ablation}
\end{table}




\subsection{Choice of Filter for Structural Anomaly Detection}
\label{sec:statistical-filter-choice}

We compare the Grubbs outlier filter with three fixed-threshold alternatives, (i) structural density ($s_d \geq 0.5$), (ii) weighted structural aggregation ($s_C \geq 0.5$), and (iii) size-gated aggregation ($s_C \geq 0.5$ and $|C|\geq10$).

Table~\ref{tab:appendix-filter-choice} shows that fixed thresholds are unstable across conversational datasets, yielding high FPR or reduced detection. In contrast, the Grubbs filter maintains high recall with substantially fewer false positives. The final setting uses the Grubbs outlier~\cite{NIST_GrubbsTest} test to select anomalous components.

\begin{table}[!htbp]
\centering
\caption{Comparison between statistical filter and threshold-based filter}
\label{tab:appendix-filter-choice}
\vspace{-4pt}
\footnotesize
\setlength{\tabcolsep}{4.5pt}
\renewcommand{\arraystretch}{0.95}
\begin{tabular}{l ccc}
\toprule
\textbf{Filter} & \textbf{FPR(\%)$\downarrow$} & \textbf{Rec(\%)$\uparrow$} & \textbf{F1(\%)$\uparrow$} \\
\midrule
\multicolumn{4}{l}{\textit{Dataset: OASST1}} \\
\midrule
$s_d \geq 0.5$                  & 34.90 & 0.47  & 1.00  \\
$s_C \geq 0.5$                  & 29.50 & 99.91 & 88.00 \\
$s_C \geq 0.5$ \& $|C|\geq 10$ & 5.58  & 100.00 & 98.00 \\
\textbf{Outlier Test}           & \textbf{0.00} & \textbf{99.91} & \textbf{100.00} \\
\midrule
\multicolumn{4}{l}{\textit{Dataset: WildChat}} \\
\midrule
$s_d \geq 0.5$                  & 25.80 & 0.47  & 1.00  \\
$s_C \geq 0.5$                  & 49.60 & 99.91 & 81.00 \\
$s_C \geq 0.5$ \& $|C|\geq 10$ & 9.50  & 100.00 & 97.00 \\
\textbf{Outlier Test}           & \textbf{6.70} & \textbf{99.91} & \textbf{97.00} \\
\midrule
\multicolumn{4}{l}{\textit{Dataset: ChatAlpaca}} \\
\midrule
$s_d \geq 0.5$                  & 46.28 & 0.38  & 1.00  \\
$s_C \geq 0.5$                  & 68.59 & 99.91 & 76.00 \\
$s_C \geq 0.5$ \& $|C|\geq 10$ & 17.69 & 99.91 & 65.00 \\
\textbf{Outlier Test}           & \textbf{3.46} & \textbf{99.72} & \textbf{98.00} \\
\bottomrule
\end{tabular}
\vspace{-1mm}
\end{table}

\section{Baseline Diversity and Hyperparameter Sensitivity}
\label{sec:hyperparameters}
\subsection{Semantic diversity of baseline}
\label{sec:baseline-diversity}

We evaluate the semantic diversity of the baseline prompt initialization using average pairwise cosine similarity between prompt embeddings, with diversity defined as one minus similarity. Prompts are deduplicated prior to analysis. 

Table~\ref{tab:base-diversity} shows high baseline diversity, with an overall score of \textbf{0.9487} and low average similarity. This confirms that the baseline is not dominated by redundant patterns, enabling reliable similarity threshold estimation with minimal bias.

\begin{table}[t]
\centering
\small
\setlength{\tabcolsep}{4pt}
\renewcommand{\arraystretch}{0.75}
\caption{Semantic diversity of baseline prompt initialization.}
\label{tab:base-diversity}
\begin{tabular}{lcc}
\toprule
\textbf{Category} & \textbf{Avg Similarity}$\downarrow$ & \textbf{Diversity}$\uparrow$ \\
\midrule
Summarization & 0.0821 & 0.9179 \\
Sentiment     & 0.3067 & 0.6933 \\
Conversation  & 0.2531 & 0.7469 \\
Reasoning     & 0.1105 & 0.8895 \\
QA            & 0.0718 & 0.9282 \\
Instruction   & 0.0426 & 0.9574 \\
\midrule
\textbf{Overall} & \textbf{0.0513} & \textbf{0.9487} \\
\bottomrule
\end{tabular}
\end{table}

\subsection{Similarity Threshold Selection}
\label{sec:threshold-selection}

We analyze the impact of the similarity threshold $\tau$ used for graph construction, which controls edge formation between queries. Table~\ref{tab:threshold_analysis} shows a clear trade-off between detection sensitivity and robustness. Higher thresholds (e.g., 90--95\%) result in conservative graphs with fewer edges, reducing recall as semantically related adversarial queries fail to connect. In contrast, lower thresholds (e.g., 60--70\%) increase connectivity, preserving recall but introducing noise that degrades F1. Performance remains stable in the range of 75--85\%, where high recall is maintained without significant loss in F1. We select $\tau=80\%$ as it achieves near-optimal recall and the highest F1, providing a robust operating point for adversarial query detection.

\begin{table}[!htbp]
\centering
\small
\setlength{\tabcolsep}{4pt}
\renewcommand{\arraystretch}{0.75}
\caption{Impact of similarity threshold on adversarial query detection performance. Relative change is computed with respect to the 80\% threshold (baseline).}
\label{tab:threshold_analysis}
\vspace{-4pt}
\scalebox{0.95}{%
\begin{tabular}{lll}
\toprule
\textbf{Threshold (\%)} & \textbf{Recall(\%)$\uparrow$} & \textbf{F1(\%)$\uparrow$} \\
\midrule
95          & 81.00 (-18.2\%)  & 90.00 (-10.0\%) \\
90          & 94.00 (-5.1\%)   & 97.00 (-3.0\%)  \\
85          & 99.00 (-0.7\%)   & 99.00 (-1.0\%)  \\
\textbf{80} & \textbf{99.00 (---)} & \textbf{100.00 (---)} \\
75          & 99.00 ($\pm$0.0\%) & 100.00 ($\pm$0.0\%) \\
70          & 100.00 (+1.0\%)  & 98.00 (-2.0\%)  \\
60          & 100.00 (+1.0\%)  & 96.00 (-4.0\%)  \\
\bottomrule
\end{tabular}
}
\end{table}


\subsection{Hyperparameter Sensitivity}
\label{sec:hyperparameters_sensitivity}

\begin{figure}[t]
  \centering
  \includegraphics[width=\columnwidth]{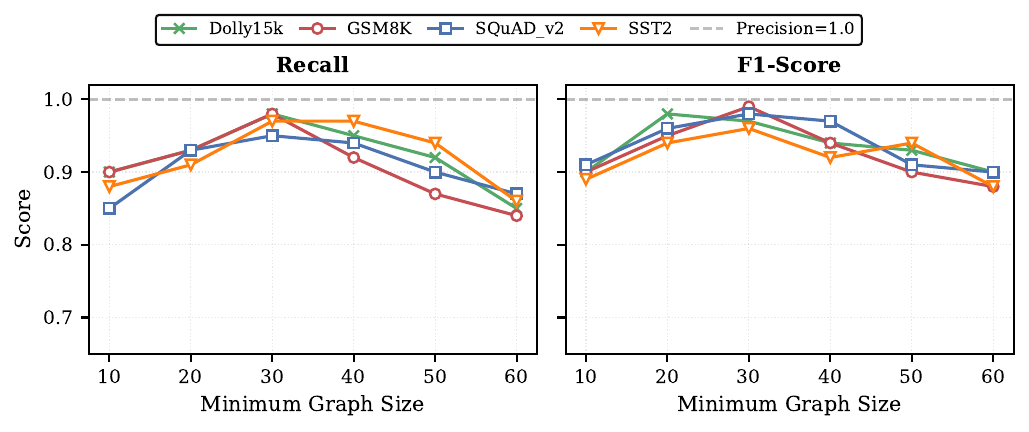}
  \caption{Impact of minimum input graph size on detection performance.}
  \label{fig:detection_performance}
\end{figure}

We evaluate the sensitivity of \defense to key hyperparameters, by varying minimum graph size ($N_{\min}$) and the detection trigger threshold (TTD), reporting mean precision, recall, and F1 over five independent trials. Figure~\ref{fig:detection_performance} averages results across datasets and seeds, and Table~\ref{tab:graph_size_sweep} provides a representative case.

\paragraph{Minimum interaction history ($N_{\min}$)}
Figure~\ref{fig:detection_performance} shows detection performance as a function of the minimum interaction history required before classification. When $N_{\min}$ is small (e.g., 10), interaction structure is insufficiently formed, leading to unstable recall and reduced F1. Increasing $N_{\min}$ to 30 stabilizes detection across all datasets, yielding near-optimal recall and F1 while maintaining perfect precision. Larger values provide no consistent benefit and introduce unnecessary delay. We therefore use $N_{\min}=30$.

\begin{figure}[t]
  \centering
  \includegraphics[width=\columnwidth]{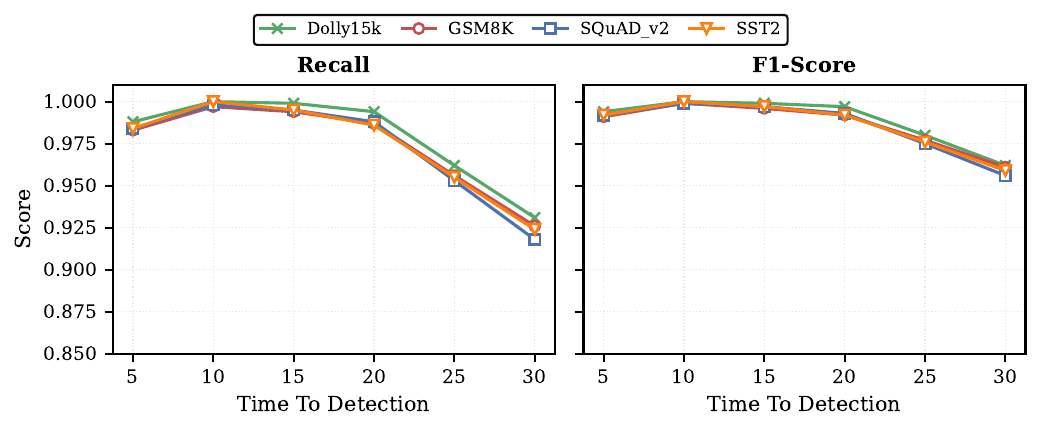}
  \caption{Effect of varying Time-to-Detect (TTD) on detection performance, evaluated at the default minimum graph size of 30 nodes.}
  \label{fig:ttd_performance}
\end{figure}

\paragraph{Time-to-detect (TTD)}
Figure~\ref{fig:ttd_performance} evaluates the effect of the TTD threshold, defined as the number of incoming queries between successive graph-based detection checks after a component reaches $N_{\min}$.
A check triggers a defense action only when the statistical filter selects the component and the GCN classifies it as malicious. Smaller values (e.g., TTD=5) enable faster detection but introduce noise, increasing false negatives. Larger values delay detection and reduce recall. Both TTD=10 and TTD=20 achieve near-perfect performance; we select TTD=10 as it provides earlier detection without sacrificing accuracy.

\begin{table}[!htbp]
\centering
\small
\caption{Impact of minimum interaction history $N_{\min}$ on detection performance. All performance metrics are reported in percentages.}
\label{tab:graph_size_sweep}
\vspace{-4pt}
\resizebox{0.9\columnwidth}{!}{
\begin{tabular}{c|rrr|ccc}
\toprule
\textbf{Graph Size} &
\textbf{TP} & \textbf{FP} & \textbf{FN} &
\textbf{FPR(\%)$\downarrow$} & \textbf{Rec(\%)$\uparrow$} & \textbf{F1(\%)$\uparrow$} \\
\midrule
5  & 600  & 25 & 726 & 4.00 & 45.20 & 61.51 \\
10 & 820  & 18 & 506 & 2.80 & 61.80 & 75.79 \\
15 & 950  & 12 & 376 & 1.90 & 71.60 & 83.04 \\
20 & 1286 & 2  & 40  & 0.40 & 96.98 & 98.39 \\
30 & 1300 & 0  & 26  & 0.00 & 98.04 & 99.00 \\
\bottomrule
\end{tabular}
}
\end{table}

\section{Runtime Overhead}
\subsection{Per-Component Latency Overhead Breakdown}
\label{sec:runtime_breakdown}

Table~\ref{tab:advguard_overhead_breakdown} reports the per-component latency of \defense. Measurements are independent and not strictly additive due to caching and overlapping execution.

\begin{table}
\centering
\footnotesize
\setlength{\tabcolsep}{3.5pt}
\renewcommand{\arraystretch}{0.90}
\caption{Per-prompt latency breakdown of \defense (ms).}
\vspace{-4pt}
\begin{tabular}{lccc}
\toprule
\textbf{Component} & \textbf{P50} & \textbf{P95} & \textbf{P99} \\
\midrule
Embedding               & 7.50  & 11.98 & 14.31 \\
Similarity (NN)         & 0.93  & 3.81  & 4.52  \\
Early-leakage           & 7.99  & 12.90 & 17.50 \\
Structural connectivity & 26.78 & 42.99 & 46.63 \\
\bottomrule
\end{tabular}
\vspace{-4pt}
\label{tab:advguard_overhead_breakdown}
\end{table}

Structural connectivity contributes the largest latency, while embedding and similarity search incur substantially lower overhead. Tail latencies follow a similar pattern, indicating predictable component-level runtime behavior.

\section{Beyond prompt leakage}
\label{sec:beyond-prompt-leakage}
\subsection{Generalization to Jailbreaking and Prompt Injection}
\label{sec:other_adversarial-attack}

\begin{table}[!htbp]
\centering
\caption{Detection performance against Jailbreaking and prompt-injection-style queries.}
\label{tab:jailbreak_prompt_injection}
\vspace{-4pt}
\begin{tabular}{@{}lccc@{}}
\toprule
\textbf{Defense} & \textbf{FPR(\%)} & \textbf{Recall(\%)} & \textbf{F1(\%)} \\
\midrule
\multicolumn{4}{@{}l}{\textit{Jailbreaking Attacks}} \\
\midrule
PromptGuard & 99.11 & 11.25 & 1.66 \\
LLMGuard & 12.70 & 93.65 & 89.30 \\ 
\textbf{AdvGuard} & \textbf{10.11} & \textbf{93.65} & \textbf{91.50} \\
\midrule
\multicolumn{4}{@{}l}{\textit{Prompt-Injection Attacks}} \\
\midrule
PromptGuard & 94.25 & 8.54 & 6.87 \\
LLMGuard & \textbf{4.31} & \textbf{94.31} & \textbf{95.58} \\
\textbf{AdvGuard} & 4.42 & 94.31 & 95.00 \\
\bottomrule
\end{tabular}
\end{table}

We evaluate whether \defense generalizes beyond prompt leakage to jailbreaking and prompt injection, using 1,050 queries from JailbreakingLLMs~\cite{chao2023jailbreaking} and 1,094 from Open Prompt Injection~\cite{liu2024promptinjection}.

Table~\ref{tab:jailbreak_prompt_injection} shows that \defense remains effective across both attacks. It achieves the lowest jailbreaking FPR (\textbf{10.11\%}) with \textbf{91.50\% F1}, and remains competitive on prompt injection with \textbf{95.00\% F1}, matching or outperforming LLMGuard while substantially outperforming PromptGuard.

\end{document}